\documentclass[aip,jcp,amsmath,amssymb,
reprint]{revtex4-2}

\usepackage[english]{babel}

\DeclareUnicodeCharacter{E5F8}{} 

\usepackage{graphicx}
\usepackage{dcolumn}
\usepackage{bm}
\usepackage{color}
\usepackage[utf8]{inputenc}
\usepackage{textcomp}
\usepackage[T1]{fontenc}
\usepackage{mathptmx}
\usepackage{etoolbox}
\usepackage{multirow}
\usepackage[normalem]{ulem}
\usepackage{adjustbox}
\usepackage[most]{tcolorbox}
\usepackage{xcolor}
\graphicspath{ {./Images/} }

\makeatletter
\def\@email#1#2{%
 \endgroup
 \patchcmd{\titleblock@produce}
  {\frontmatter@RRAPformat}
 {\frontmatter@RRAPformat{\produce@RRAP{*#1\href{mailto:#2}{#2}}}\frontmatter@RRAPformat}
  {}{}
}%
\makeatother

\begin{document}

\title[Onset of Stripe Order]{Onset of Stripe Order in Classical Fluids: Lessons from Lattice-Gas Mixtures}

\author{G. Costa}
\author{S. Prestipino}
\email{gabriele.costa@studenti.unime.it, sprestipino@unime.it}
\affiliation{Dipartimento di Scienze Matematiche e Informatiche, Scienze Fisiche e Scienze della Terra, Università degli Studi di Messina, Viale F. Stagno d’Alcontres 31, 98166 Messina, Italy}

\begin{abstract}
When two molecular species with mutual affinity are mixed together, various self-assembled phases can arise at low temperature, depending on the shape of like and unlike interactions.
Among them, stripes --- where layers of one type are regularly alternated with layers of another type --- hold a prominent place in materials science, occurring e.g.
in the structure of superconductive doped antiferromagnets.
Stripe patterns are relevant for the design of functional materials, with applications in optoelectronics, sensing, and biomedicine.
In a purely classical setting, an open question pertains to the features that spherically-symmetric particle interactions must have to foster stripe order.
Here we address this challenge for a lattice-gas mixture of two particle species, whose equilibrium properties are exactly determined by Monte Carlo simulations with Wang-Landau sampling, in both planar and spherical geometry, and for equal chemical potentials of the species.
Somewhat surprisingly, stripes can emerge from largely different off-core interactions, featuring various combinations of repulsive like interactions with a predominantly attractive unlike interaction.
In addition to stripes, our survey also unveils crystals and crystal-like structures, cluster crystals, and networks, which considerably broaden the catalog of possible patterns.
Overall, our study demonstrates that stripes are more widespread than generally thought, as they can be generated by several distinct mechanisms, thereby explaining why stripe patterns are observed in systems as diverse as cuprate materials, biomaterials, and nanoparticle films.
\end{abstract}

\maketitle

\section{Introduction}
\begin{figure}
\includegraphics[width=6cm]{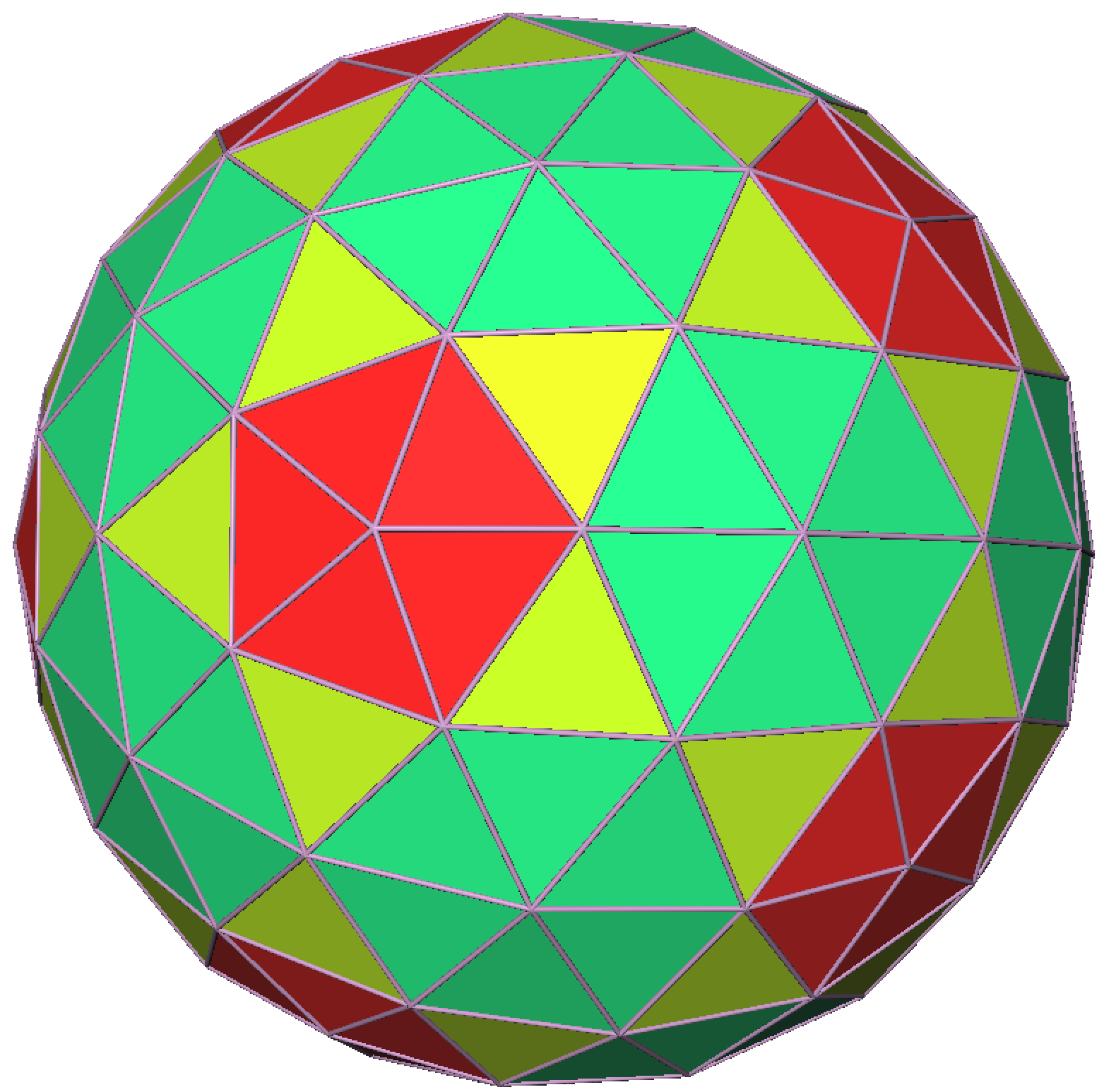}
\caption{\label{HPCD}The polyhedron depicted in the figure is the HPCD.
It is a geodesic icosahedron with 240 triangular faces and 122 vertices, of which twelve only are fivefold vertices;
the other vertices are sixfold coordinated.
Different colors are used for different types of triangles.}
\end{figure}

In the field of condensed matter physics, the interplay between microscopic interactions and spatial constraints can lead to complex patterns across multiple scales~\cite{meinhardt2003models,hoyle2006pattern,seul1995domain}.
Among them, stripes (i.e., a regular alternation of two kinds of layers) stand out as one of the most notable arrangements, and the central motif in a wide variety of material systems, with applications in areas like optoelectronics and sensing.
Stripe patterns are found in systems as diverse as nickelates~\cite{tranquada1994simultaneous}, cuprate superconductors~\cite{tranquada1996neutron,wu2011magnetic,vojta2009lattice}, and ultracold gases~\cite{wang2023evidence}.
In these strongly-correlated materials, stripe formation is grounded on the existence of competing interactions, which drive electrons/atoms toward states with spontaneously broken spatial symmetry.
Self-assembly into spatially modulated structures --- usually referred to as microphases --- is a recurrent phenomenon and a key area of study also in soft matter~\cite{zhuang2016recent,zhuang2016b}, and is as well governed by counteracting forces:
a short-range attraction (SA), often arising from depletion forces, which promotes aggregation, and a long-ranged repulsion (LR), due to screened electrostatic forces, which inhibits macroscopic phase separation.
SALR systems can exhibit several microphases, depending on the interaction parameters and thermodynamic conditions~\cite{ciach2013origin,pini2017}:
clusters and reverse clusters, lamellae/stripes, and gyroid phases (to name just the prominent ones).
Microphases are found in protein solutions~\cite{stradner2004equilibrium}, diblock copolymers~\cite{bates201750th,bates1999block}, and colloidal dispersions~\cite{tierno2009colloidal,mino2012colloidal,das2018formation}.
Given the relevance of stripes to both fundamental science and technology, a key question is how peculiar microscopic interactions should be to favor stripe ordering.
Addressing this issue is of the utmost importance for achieving a better control over the onset of stripes, and this is where theoretical modeling and numerical simulations can play a crucial role.
In one-component systems of classical particles, the requirements for observing stripes are relatively well understood.
Notably, aside from fluids with competing interactions~\cite{imperio2006microphase,archer2008two,pekalski2014periodic}, stripes are also found in systems governed by a purely repulsive potential~\cite{malescio2003stripe,malescio2004stripe,pattabhiraman2017formation}.
However, other situations possibly exist that are propitious to the appearance of stripes.
For example, we have recently demonstrated for systems of particles on a spherical grid that even Lennard-Jones-type interactions can stabilize worm-like arrangements at low temperature~\cite{costa2025b}.
Stable stripe phases have been reported even on a quasicrystalline lattice~\cite{teixeira2025}.
In recent years, the investigation of stripes in binary mixtures of classical particles has been attracting growing attention.
Due to a vast number of possible combinations, binary mixtures are still relatively little explored and only scattered results are available~\cite{mendoza2009self,ciach2011,patsahan2021,padilla2021dynamics,litniewski2024adsorption,litniewski2025};
therefore, it is fair to say that the question about the origin and variety of stripe order in mixtures is so far unanswered in full generality.
Clearly, the addition of a second species significantly alters self-assembly.
For example, a mixture of hard spheres and SALR particles, interacting through a square-well (SW) cross attraction, has been shown to exhibit clusters under thermodynamic conditions that would not allow for their existence in the pure SALR fluid~\cite{munao2022competition}.
Species asymmetry is not even necessary for the emergence of microphases.
Indeed, a non-trivial self-assembly (including stripes) has been observed in mixtures of two equivalent species characterized by a repulsive (i.e., hard-sphere plus shoulder) like interaction and a SW unlike attraction~\cite{mendoza2009self,padilla2021dynamics}.
In Refs.~\onlinecite{munao2023like,prestipino2023density,Prestipino2025} like interactions are even simpler, i.e., of purely hard-core type, and yet the phase behavior still features various kinds of stripes.
These findings suggest that the occurrence of stripes is essentially governed by the range of cross attraction, rather than by the details of like interactions.
A model mixture involving a SALR potential between like particles and an exactly opposite potential between particles of a different species has been investigated both in the continuum~\cite{patsahan2024spontaneous} and in a lattice framework~\cite{ciach2023pattern,devirgiliis2024,devirgiliis2024lattice};
these studies revealed that the emergence and stabilization of ordered patterns, including stripes and clusters, are highly sensitive to the repulsion-to-attraction ratio in the SALR potential, regardless of the nature of the hosting space.
In two recent papers~\cite{costa2025,costa2025b}, we have investigated several one-component lattice gases defined on spherical grids constructed from the vertices of a geodesic icosahedron~\cite{visualpolyhedra} (i.e., a semiregular polyhedron with triangular faces and the least possible number of fivefold vertices), finding, despite geometric frustration, a wide variety of low-temperature ``phases'', including regular polyhedra, cluster crystals, and worm-like patterns.
In the present paper, the same investigation is extended to binary mixtures, now choosing as ambient space the grid formed by the vertices of a hexakis-pentakis chamfered dodecahedron (HPCD, see Fig.~\ref{HPCD}).
The HPCD is a geodesic icosahedron with sufficiently many vertices (122) to produce an emergent (i.e., thermodynamic) behavior.
Clearly, the geometry of the grid impacts on the nature of the emergent structures, since it affects particle coordination and, consequently, the energy of individual configurations.
However, for a comparison with the triangular lattice the most natural choice is by far a geodesic grid, since it allows us to keep the contact with the planar model as closest as possible.
Even though a curved grid is generally detrimental to the onset of stripe order, the tendency to forming stripes would be apparent, e.g., in the alternation of rows/ribbons of particles of same species, wrapped around the sphere (as found, e.g., in Ref.~\onlinecite{perez2022packing}).
In another respect, a spherical grid might accommodate stripe-like structures of a type unknown to a planar lattice.
To assess the influence of curvature, the lowest-energy arrangements on the spherical grid will be compared with the structures promoted by the same interaction on the triangular lattice.
Clearly, a comprehensive study of self-assembly involving several different interactions can only be accomplished if we could rely on a fast and accurate method to extract the system ``phases'' from the Hamiltonian.
To this end, we will employ Wang-Landau sampling~\cite{wang2001efficient,wang2001determining,landau2004new}, which already has proved effective in the study of one-component systems~\cite{costa2025,costa2025b}.
Investigations of particles confined to a spherical surface are not a novelty:
many studies have addressed self-assembly on a sphere, both in one-component fluids~\cite{prestipino1993,zarragoicoechea2009pattern,pekalski2018orientational,franzini2022phase,ciardi2024supersolid,perez2022packing} and in binary mixtures~\cite{meyra2010monte,dlamini2021self,zhao2021self,litniewski2024adsorption}.
In particular, particle arrangements with icosahedral symmetry have been observed in several cases~\cite{bowick2000interacting,de2015entropy,paquay2016energetically,prestipino2019}, including the proteins making up a virus capsid~\cite{caspar1962physical,zandi2004origin}.
However, as far as we know, no previous study has examined the equilibrium behavior of binary mixtures on a spherical grid.
More importantly, in none of the aforementioned papers a broad range of like and unlike interactions has been explored for a systematic search of stripe patterns.
We believe that neither a lower dimensionality nor space discretization would sensibly affect the generality of our considerations about the nature of the interactions responsible for the emergence of stripes in realistic mixtures.
The paper is organized as follows.
In Section II we introduce the model system and describe the method used to extract its statistical properties.
The results obtained are presented and commented in Section III-V.
The final Section VI is devoted to conclusions and outlook.

\section{Model and method}

We investigate binary mixtures of non-overlapping particles, living on a finite grid of $M$ sites.
Particles interact through a Hamiltonian $H[c]$, where $c=\{c_{1,1},c_{1,2},\ldots,c_{M,1},c_{M,2}\}$ is the generic system microstate, labeled by the occupation numbers (0 or 1) of the sites (for example, $c_{i,1}=1$ if the $i$-th site is occupied by a particle of species 1; in this case, $c_{i,2}=0$).
Single site occupancy prevents the possibility of a Ruelle-type instability~\cite{Ruelle1969}.
The general framework adopted in the present paper is thus similar to Ref.~\onlinecite{costa2025b}, but only one curved grid is considered here, namely the spherical grid obtained by projecting the vertices of the HPCD polyhedron onto a reference sphere ($M=122$).
The same mixtures are also examined on a $12\times 12$ triangular-lattice grid, which is periodically replicated in every direction in order to reduce finite-size effects.
To simplify our study, the sites of the spherical grid are gathered in coordination shells, and all sites in the same shell are assigned the same interaction energy with the central site~\cite{costa2025b};
furthermore, no interaction occurs beyond the third shell.
Denoting $u^{(n)}_{\alpha,\beta}$ (with $n=1,2,3$ and $\alpha,\beta=1,2$) the strength of the interaction between two particles of species $\alpha$ and $\beta$ that are $n$-th neighbor of each other, the Hamiltonian of the mixture reads:
\begin{eqnarray}
H[c]&=&\sum_{\rm 1NP}\sum_{\alpha,\beta}u^{(1)}_{\alpha,\beta}c_{i,\alpha}c_{j,\beta}+\sum_{\rm 2NP}\sum_{\alpha,\beta}u^{(2)}_{\alpha,\beta}c_{i,\alpha}c_{j,\beta}
\nonumber \\
&+&\sum_{\rm 3NP}\sum_{\alpha,\beta}u^{(3)}_{\alpha,\beta}c_{i,\alpha}c_{j,\beta}\,,
\label{eq1}
\end{eqnarray}
where the first sum runs over all distinct first-neighbor pairs (1NP) of sites, and so on.
For each $n$ we distinguish two like interactions ($u^{(n)}_{1,1}$ and $u^{(n)}_{2,2}$) and one unlike interaction ($u^{(n)}_{1,2}=u^{(n)}_{2,1}$).
A lattice-gas mixture is then characterized by a total of nine coupling parameters;
only if the species are of same nature (``symmetric mixture''), then $u^{(n)}_{1,1}=u^{(n)}_{2,2}$.
Since particles occupy discrete positions on the sphere, interactions of range shorter than the minimum separation between two grid points cannot be considered in our setup.
By an appropriate choice of couplings, it is possible to mimic various combinations of like and unlike interactions in two dimensions, including core-corona, Lennard-Jones, and SALR, allowing us to compare their thermal properties and structural motifs.
Specifically, any repulsive interaction characterized by non-negative couplings of decreasing strength is of core-corona-type, while we designate Lennard-Jones-type any interaction exhibiting an attractive ``well'', possibly preceded by a repulsive ``shoulder''.
Finally, the interaction is classified SALR when a short-range attraction is followed by a longer-range repulsion.
Hereafter, all couplings are expressed in units of a positive energy $\epsilon$.
Accordingly, the temperature $T$ is given in units of $\epsilon/k_{\rm B}$, where $k_{\rm B}$ is Boltzmann’s constant.
Once the couplings have been chosen, the statistical properties of the mixture are evaluated in the grand-canonical ensemble using the Monte Carlo (MC) method with Wang–Landau (WL) sampling.
The MCWL method is particularly suited for discrete systems with not too many degrees of freedom, and is especially effective at low temperature where conventional sampling techniques often fail due to the presence of multiple free-energy minima separated by high barriers.
The MCWL method virtually enables computing the exact density of states $g_{{\cal N}_1,{\cal N}_2,{\cal E}}$, where ${\cal N}_1$ and ${\cal N}_2$ are the number of type-1 and type-2 particles, respectively, and ${\cal E}$ is the total energy.
The reader is referred to our previous paper~\cite{costa2025b} for a description of the algorithm.
In practice, in order to reduce the computational burden, we arbitrarily decide to study only mixtures of two species with same chemical potential $\mu$.
By this choice, which gives easy access to the total number density as a function of temperature and chemical potential, the patterns developed by a symmetric mixture would all be near-equimolar, at least unless the unlike interaction is repulsive.
We ensure a thorough exploration of the mixture state space by considering three types of MC moves:
besides the usual creation and annihilation moves, other moves are attempted that change the particle species (from 1 to 2, or vice versa).
We also study equimolar mixtures (${\cal N}_1={\cal N}_2$).
Then, depending on the occupancies of two randomly chosen sites, we either create/destroy two unlike particles or swap the species of the particles;
if none of these cases apply, two new sites are chosen and a new move is attempted.
Trial moves are accepted with the MCWL acceptance criterion, involving the current density of states at old and new particle-number and energy values~\cite{wang2001efficient,wang2001determining,landau2004new}.
Given the density of states, the grand-canonical partition function for $\mu_1=\mu_2=\mu$ immediately follows as
\begin{equation}
\Xi(T,\mu)=\sum_{{\cal N},{\cal E}}{\cal G}_{{\cal N},{\cal E}}\,e^{\beta\mu{\cal N}}e^{-\beta{\cal E}}\,,
\label{eq2}
\end{equation}
where $\beta=(k_{\rm B}T)^{-1}$ and ${\cal G}_{{\cal N},{\cal E}}=\sum_{{\cal N}_1,{\cal N}_2}^{({\cal N})}g_{{\cal N}_1,{\cal N}_2,{\cal E}}$.
The latter sum is over all $({\cal N}_1,{\cal N}_2)$ pairs such that ${\cal N}_1+{\cal N}_2={\cal N}$.
In the equimolar case, the partition function is similarly expressed by Eq.~\ref{eq2} with $\mu=(\mu_1+\mu_2)/2$, but ${\cal N}$ now runs over even numbers only and ${\cal G}_{{\cal N},{\cal E}}=g_{\frac{\cal N}{2},\frac{\cal N}{2},{\cal E}}$.
The grand potential is $\Omega=-k_{\rm B}T\ln\Xi$ and the pressure is $P(T,\mu)=-\Omega/M=(k_{\rm B}T/M)\ln\Xi$.
Finally, the thermodynamic values of the total number of particles and total energy are respectively computed as:
\begin{eqnarray}
N=\langle{\cal N}\rangle=\frac{\sum_{{\cal N},{\cal E}}{\cal N}{\cal G}_{{\cal N},{\cal E}}\,e^{\beta\mu{\cal N}}e^{-\beta{\cal E}}}{\Xi}\,;
\nonumber \\
E=\langle{\cal E}\rangle=\frac{\sum_{{\cal N},{\cal E}}{\cal E}{\cal G}_{{\cal N},{\cal E}}\,e^{\beta\mu{\cal N}}e^{-\beta{\cal E}}}{\Xi}\,.
\label{eq3}
\end{eqnarray}
From the expressions of $P$ and $N$, it easily follows that
\begin{equation}
\frac{\partial P}{\partial\mu}\biggr\vert_T=\frac{N}{M}\equiv\rho\,,
\label{eq4}
\end{equation}
where $\rho$ is the number density.
At fixed temperature, $N(\mu)$ is a monotonically increasing function, which for $T\ll 1$ exhibits a number of regions where $N$ is nearly constant (``plateaus'', associated with distinct ``phases''), with steep crossovers between adjacent regions (``phase transitions'').
Related plateaus and crossovers are found in $E(\mu)$ too.
While true phase transitions are forbidden in finite systems, a plateau in $N(\mu)$ indicates that the system is structurally robust against small changes in the chemical potential.
Therefore, each plateau corresponds to a specific self-assembled structure or {\em pattern}.
Since the organizing principle at fixed $T$ and $\mu$ is the minimization of the generalized grand potential~\cite{costa2025b}, a strict plateau at $T=0$ implies that in an interval of $\mu$ the stable structure is provided by the single configuration (up to rotations and reflections, and unless accidental degeneracy occurs) that minimizes the total energy for the given $N$.
However, it is clear that well-defined patterns are exclusive of low $T$ only;
as thermal fluctuations grow stronger, finite-size effects eventually take over and the density/energy function becomes increasingly smooth.

\begin{table*}
\caption{\label{tab:prova} For each symmetric interaction (columns 2 and 3), we shortly describe the self-assembled structures of the mixture on a $12$x$12$ triangular grid (column 4).
For convenience, the various cases are marked with a progressive number (column 1).
An asterisk besides the label indicates that stripes are among the stable phases.}
\begin{tabular}{|l|c|c|c|}
\hline
\textbf{Case}&\textbf{$U_{\rm 11}=U_{\rm 22}$}& \textbf{$U_{\rm 12}$} & \textbf{Low-temperature structures} \\
\hline
$1$&$(0,0,0)$ & $(1,0,0)$ & Small $N$: Polydisperse cluster fluid.
Large $N$: One species only\\
\hline
$2$&$(0,0,0)$ & $(1,1,0)$ & Small $N$: Polydisperse cluster fluid.
Large $N$: One species only\\
\hline
$3$&$(1,0,0)$&$(0,0,0)$& Moderately large $N$: Compositionally-ordered crystal. $N=144$: Linear arrangements of like particles\\
\hline
$4*$&$(1,1,0)$&$(0,0,0)$& Intermediate $N$: Two-color stripes.
$N=144$: Stripes\\
\hline
$5*$&$(1,1,1)$&$(0,0,0)$& Moderately small $N$: Polydisperse cluster fluid. Intermediate $N$: Mix of clusters and curved stripes.\\&&& Moderately large $N$: Coexistence of different stripes.
$N=144$: Stripes\\
\hline
$6$&$(0,0,0)$ & $(-1,0,0)$ & Vapor coexisting with a compositionally-disordered solid\\
\hline
$7*$&$(0,0,0)$ & $(-1,-1,0)$ & Vapor coexisting with a stripe solid\\
\hline
$8*$&$(0,0,0)$ & $(-1,-1,-1)$ & Vapor coexisting with a stripe solid\\
\hline
$9*$&$(0,0,0)$ & $(1,-1,0)$ & Moderately small $N$: compositionally-disordered crystal.\\&&& Moderately large $N$: Stripes with ordered voids.
$N=144$: Stripes\\
\hline
$10*$&$(0,0,0)$ & $(1,0,-1)$ & Intermediate $N$: Stripes of alternating ``color''. $N=144$: Irregular stripes\\
\hline
$11*$&$(0,0,0)$& $(-1,2,1)$ & Intermediate $N$: Stripes.
$N=144$: triangular crystal\\
\hline
$12$&$(0,0,0)$ & $(-1,1,0)$ & Vapor coexisting with a triangular crystal\\
\hline
$13*$&$(0,0,0)$ & $(-1,0,1)$ & Vapor coexisting with a stripe solid\\
\hline
$14$& $(1,0,0)$& $(-1,0,0)$ & Moderately large $N$: Compositionally-ordered crystal.
$N=144$: Linear arrangements of like particles\\
\hline
$15*$& $(1,0,0)$& $(-1,-1,0)$ & Vapor coexisting with a stripe solid\\
\hline
$16*$& $(1,0,0)$& $(-1,-1,-1)$ & Vapor coexisting with a solid of irregular stripes\\
\hline
$17$& $(1,0,0)$& $(1,-1,0)$ & Moderately small $N$: compositionally-disordered solid.\\&&& Moderately large $N$: Disordered fluid with ordered voids.
Large $N$: Compositionally-disordered solid\\
\hline
$18*$& $(1,0,0)$& $(1,0,-1)$ & Moderately small $N$: compositionally-disordered crystal.
Intermediate $N$: Compositionally-disordered stripes.\\&&& Moderately large $N$: Linear arrangements of like particles with ordered voids.
Large $N$: Defective stripe solid\\
\hline
$19*$&$(1,0,0)$& $(-1,2,1)$ & Moderately small $N$: Compositionally-ordered network. Intermediate $N$: Stripes.
$N=144$: Triangular crystal\\
\hline
$20$&$(1,0,0)$ & $(-1,1,0)$ & Moderately large $N$: Compositionally-ordered crystal. $N=144$: Triangular crystal\\
\hline
$21*$&$(1,0,0)$&$(-1,0,1)$ & Moderately small $N$: Compositionally-ordered network.
$N=144$: Stripes\\
\hline
$22*$& $(1,1,0)$ & $(-1,0,0)$ & Intermediate $N$: Stripes of alternating ``color''.
$N=144$: Stripes\\
\hline
$23*$& $(1,1,0)$& $(-1,-1,0)$ & Vapor coexisting with a stripe solid\\
\hline
$24*$& $(1,1,0)$&$(1,-1,0)$& Small $N$: Compositionally-ordered crystal.
Moderately small $N$: Compositionally-disordered crystal.\\&&& Intermediate $N$: Stripes. $N=144$: Compositionally-disordered crystal\\
\hline
$25*$& $(1,1,0)$&$(1,0,-1)$& Small $N$: Compositionally-disordered crystal.
Intermediate $N$: Compositionally-disordered stripes.\\&&&  Moderately large $N$: Linear arrangements of like particles with ordered voids.
$N=144$: irregular stripes\\
\hline
$26*$& $(1,1,0)$&$(-1,2,1)$&  Small $N$: Two-color stripes. Intermediate and moderately large $N$: Stripes.\\&&& $N=144$: Stripes and triangular crystal\\
\hline
$27*$& $(1,1,1)$& $(-1,0,0)$ & Small $N$: Cluster crystal.
Intermediate $N$: irregular stripes. $N=144$: irregular stripes\\
\hline
$28*$& $(1,1,1)$& $(-1,-1,0)$ & Small $N$: Compositionally-ordered crystal.
Intermediate $N$: Two-color stripes.\\&&&  $N=144$: Stripes\\
\hline
$29$& $(1,1,1)$& $(-1,2,1)$ &Small $N$: Cluster crystal.
Moderately small $N$: Irregular stripes.\\&&& Intermediate $N$: Stripes. $N=144$ triangular crystal\\
\hline
$30$&$(-1,0,0)$ & $(-1,0,0)$ &Vapor coexisting with a compositionally-disordered solid \\
\hline
$31$&$(-1,0,0)$ & $(-1,2,1)$ & Vapor coexisting with a one-species solid\\
\hline
$32$&$(1,-1,0)$ & $(-1,0,0)$ & Moderately large $N$: Compositionally-ordered crystal.
$N=144$: Triangular crystal\\
\hline
$33*$&$(1,0,-1)$ & $(-1,0,0)$ & Vapor coexisting with a stripe solid.
The nature of stripes depend on $N$\\
\hline
$34*$&$(-1,2,1)$ &$(1,0,0)$& Moderately small and intermediate $N$: Cluster crystal.
$N=144$: Stripes \\
\hline
$35*$&$(-1,2,1)$ &$(1,1,0)$& Small $N$: Cluster fluid. Moderately small $N$: Cluster crystal.\\&&& Intermediate and moderately large $N$: Two-color stripes.
$N=144$: Stripes\\
\hline
$36*$&$(-1,2,1)$ &$(1,1,1)$& Small $N$: Cluster fluid. Moderately small $N$: Cluster crystal.
Intermediate and moderately large $N$: Stripes.\\&&& $N=144$: Stripes\\
\hline
$37*$&$(-1,2,1)$&$(-1,0,0)$& Moderately large $N$: Coexistence of different stripes.
$N=144$: Irregular stripes\\
\hline
$38*$&$(-1,2,1)$& $(1,-1,0)$ & Small $N$: Cluster fluid. Intermediate and moderately large $N$: Stripes.\\&&& $N=144$: Stripes.\\
\hline
$39*$&$(-1,1,0)$ & $(1,-1,0)$ & Vapor coexisting with stripe solid.
The nature of two-phase states depend on $N$\\
\hline
$40*$&$(-1,0,1)$ & $(1,0,-1)$ & Moderately large $N$: Stripes with ordered voids.
$N=144$: Irregular stripes\\
\hline
\end{tabular}
\end{table*}

\begin{figure*}[htbp]
  \centering
  \begin{tcolorbox}[colback=gray!20, colframe=gray!20, boxrule=0pt,
                    arc=2mm, left=1mm, right=1mm, top=1mm, bottom=1mm, width=\textwidth]
    \centering
    \begin{minipage}[t]{0.23\textwidth}
      \centering\includegraphics[height=3cm]{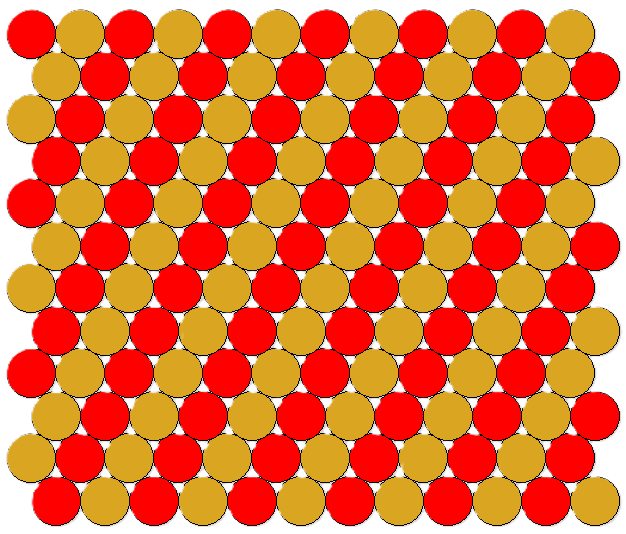}\\
      \centering \# 1 (4,7,13,15,21,22,23,33)
    \end{minipage}\hfill
    \begin{minipage}[t]{0.23\textwidth}
      \centering\includegraphics[height=3cm]{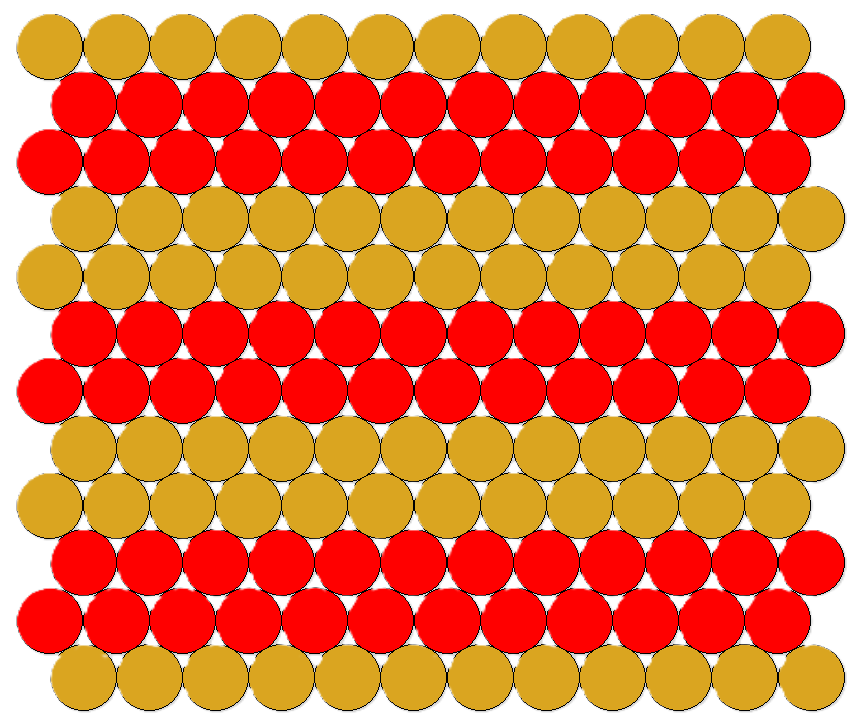}\\
      \centering \# 2 (9,34,35,38,39)
    \end{minipage}\hfill
    \begin{minipage}[t]{0.23\textwidth}
      \centering\includegraphics[height=3cm]{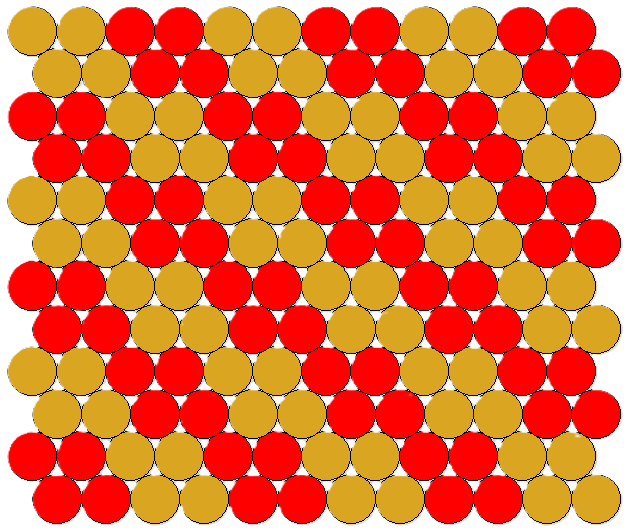}\\
      \centering \# 3 (5,8,28)
    \end{minipage}\hfill
    \begin{minipage}[t]{0.23\textwidth}
      \centering\includegraphics[height=3cm]{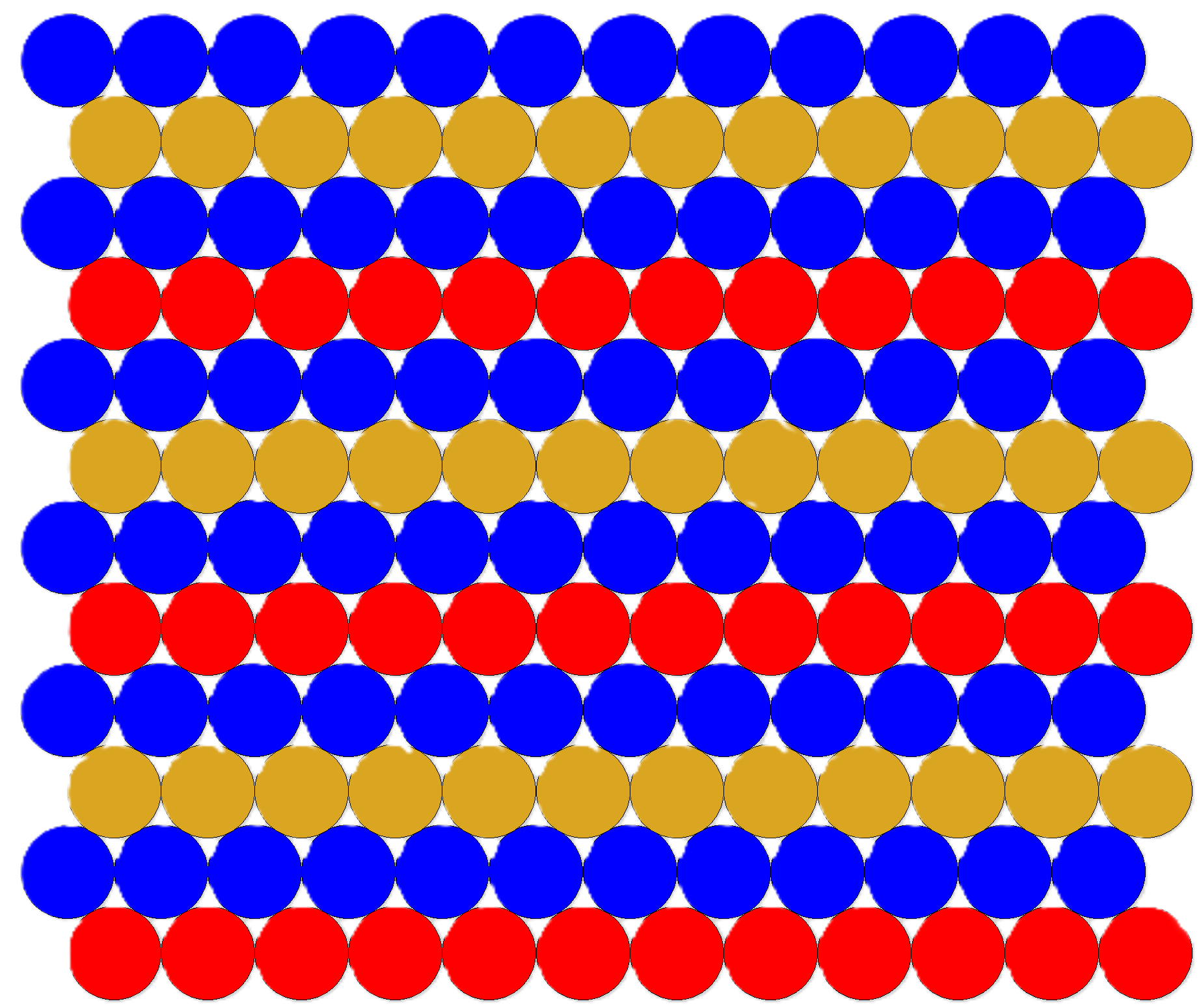}\\
      \centering \# 4 (10)
    \end{minipage}
  \end{tcolorbox}
  
  \vspace{0.5em}

  \noindent
  \begin{minipage}[b]{0.74\textwidth} 
    \begin{tcolorbox}[colback=gray!20, colframe=gray!20, boxrule=0pt,
                      arc=2mm, left=1mm, right=1mm, top=1mm, bottom=1mm, width=\linewidth]
      \centering
      \begin{minipage}[t]{0.32\linewidth}
        \centering\includegraphics[height=3cm]{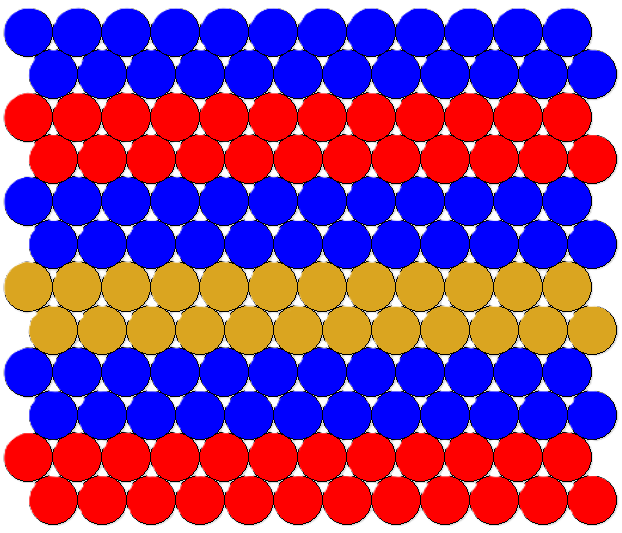}\\
        \centering \# 5 (36)
      \end{minipage}\hfill
      \begin{minipage}[t]{0.32\linewidth}
        \centering\includegraphics[height=3cm]{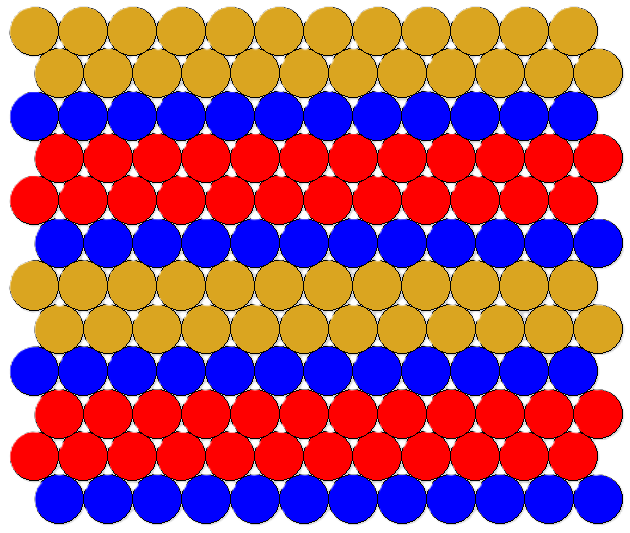}\\
        \centering \# 6 (35,36)
      \end{minipage}\hfill
      \begin{minipage}[t]{0.32\linewidth}
        \centering\includegraphics[height=3cm]{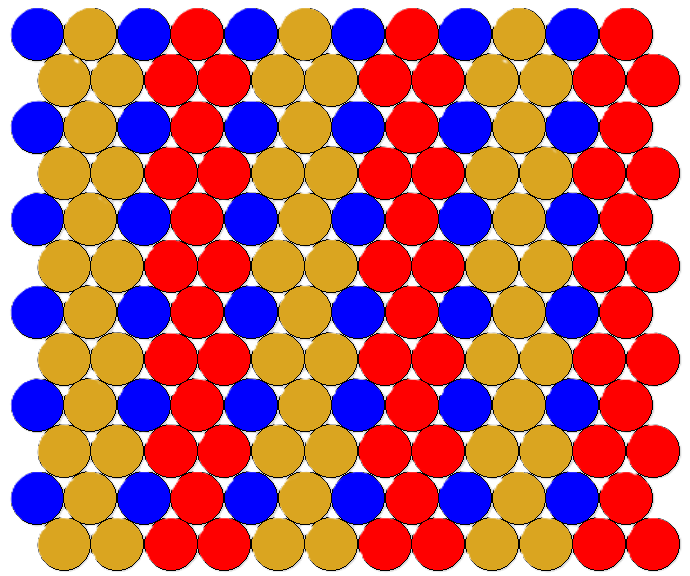}\\
        \centering \# 7 (40)
      \end{minipage}
    \end{tcolorbox}
  \end{minipage}%
  \hfill 
  \begin{minipage}[b]{0.24\textwidth} 
    \centering
    \includegraphics[height=3cm]{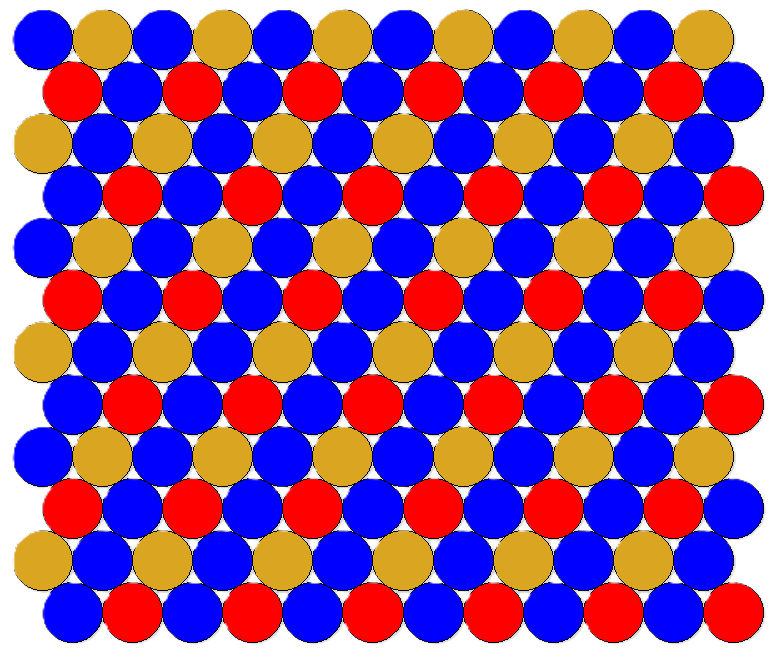}\\
    \centering \# 8 (4,22)
    \vspace{4mm} 
  \end{minipage}

  \vspace{1em}

  \begin{minipage}[t]{0.24\textwidth}
    \centering\includegraphics[height=3cm]{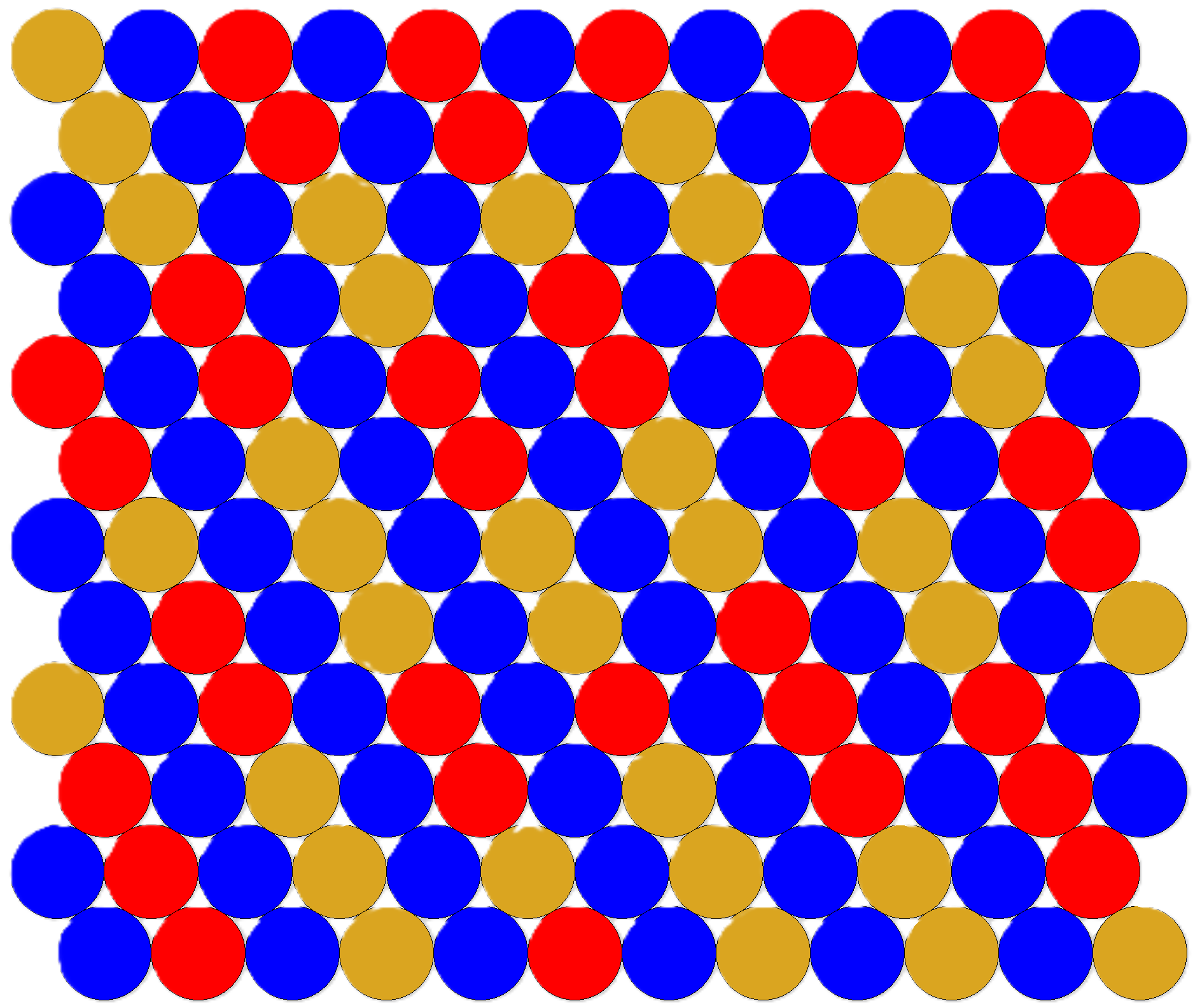}\\
    \centering \# 9 (18)
  \end{minipage}\hfill
  \begin{minipage}[t]{0.24\textwidth}
    \centering\includegraphics[height=3cm]{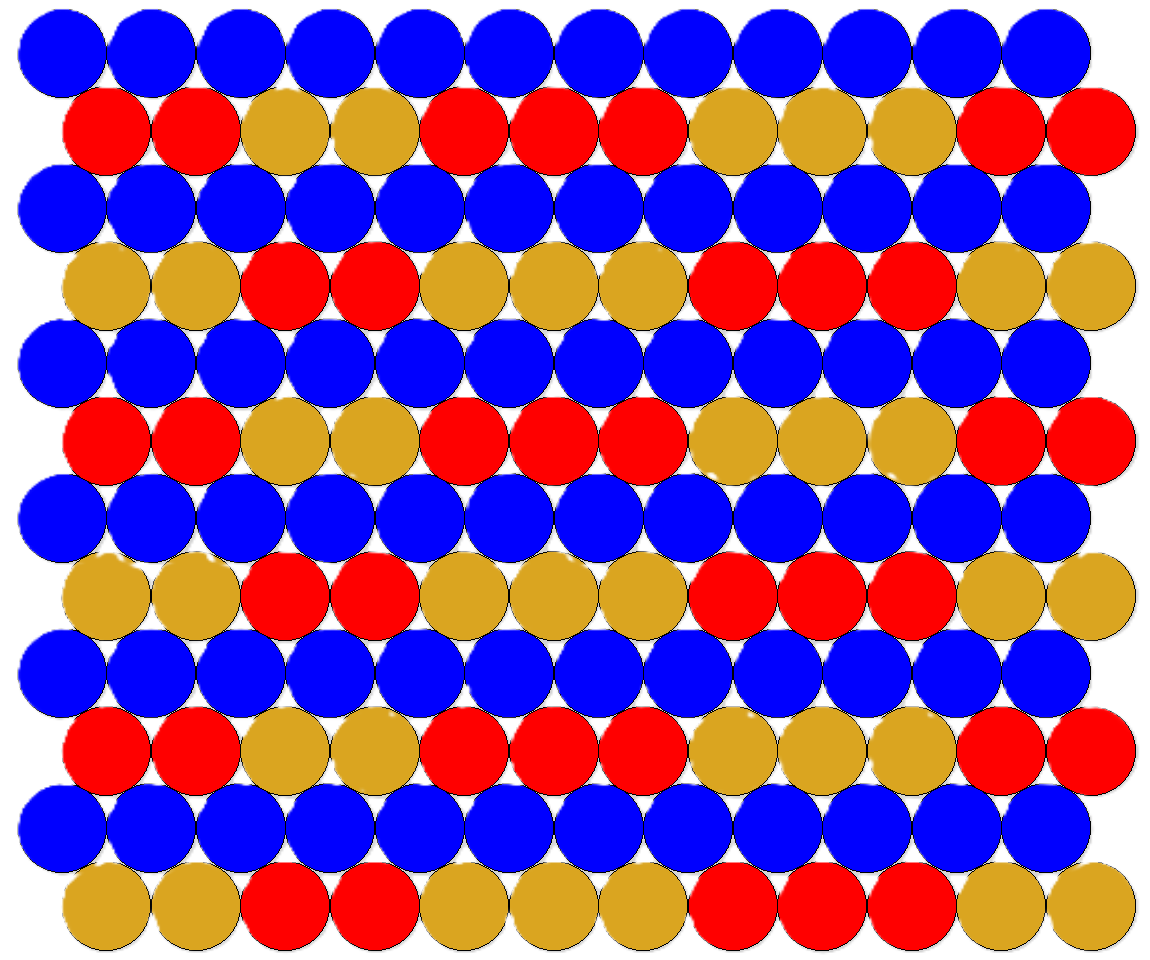}\\
    \centering \# 10 (25)
  \end{minipage}\hfill
  \begin{minipage}[t]{0.24\textwidth}
    \centering\includegraphics[height=3cm]{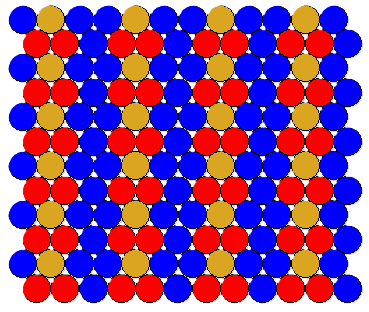}\\
    \centering \# 11 (11,19,26,29)
  \end{minipage}\hfill
  \begin{minipage}[t]{0.24\textwidth}
    \centering\includegraphics[height=3cm]{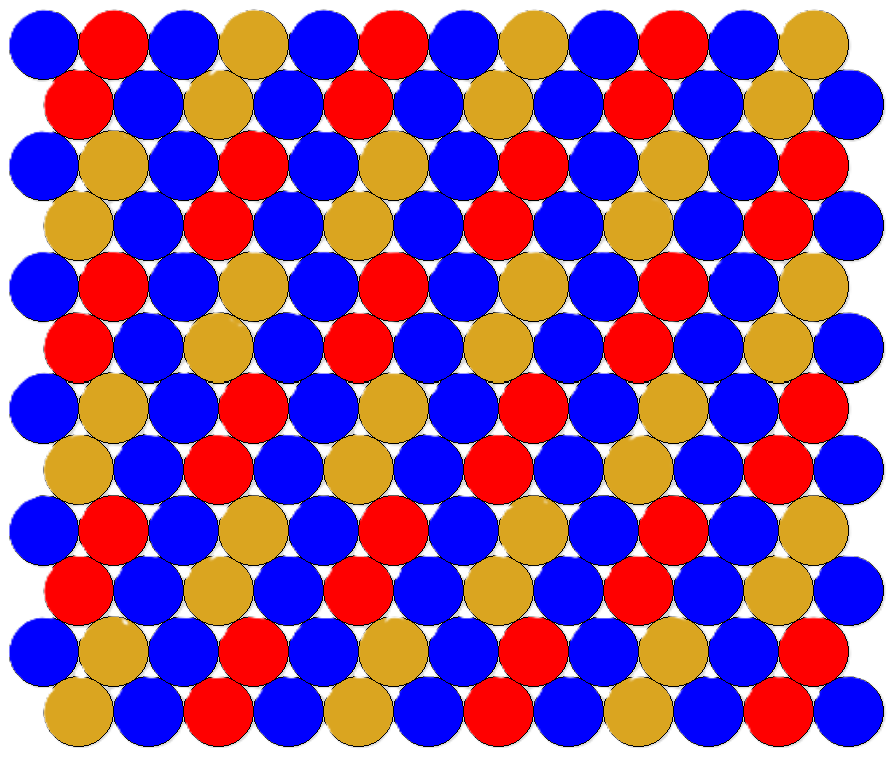}\\
    \centering \# 12 (28)
  \end{minipage}

  \vspace{1em}

  \begin{minipage}[t]{0.24\textwidth}
    \centering\includegraphics[height=3cm]{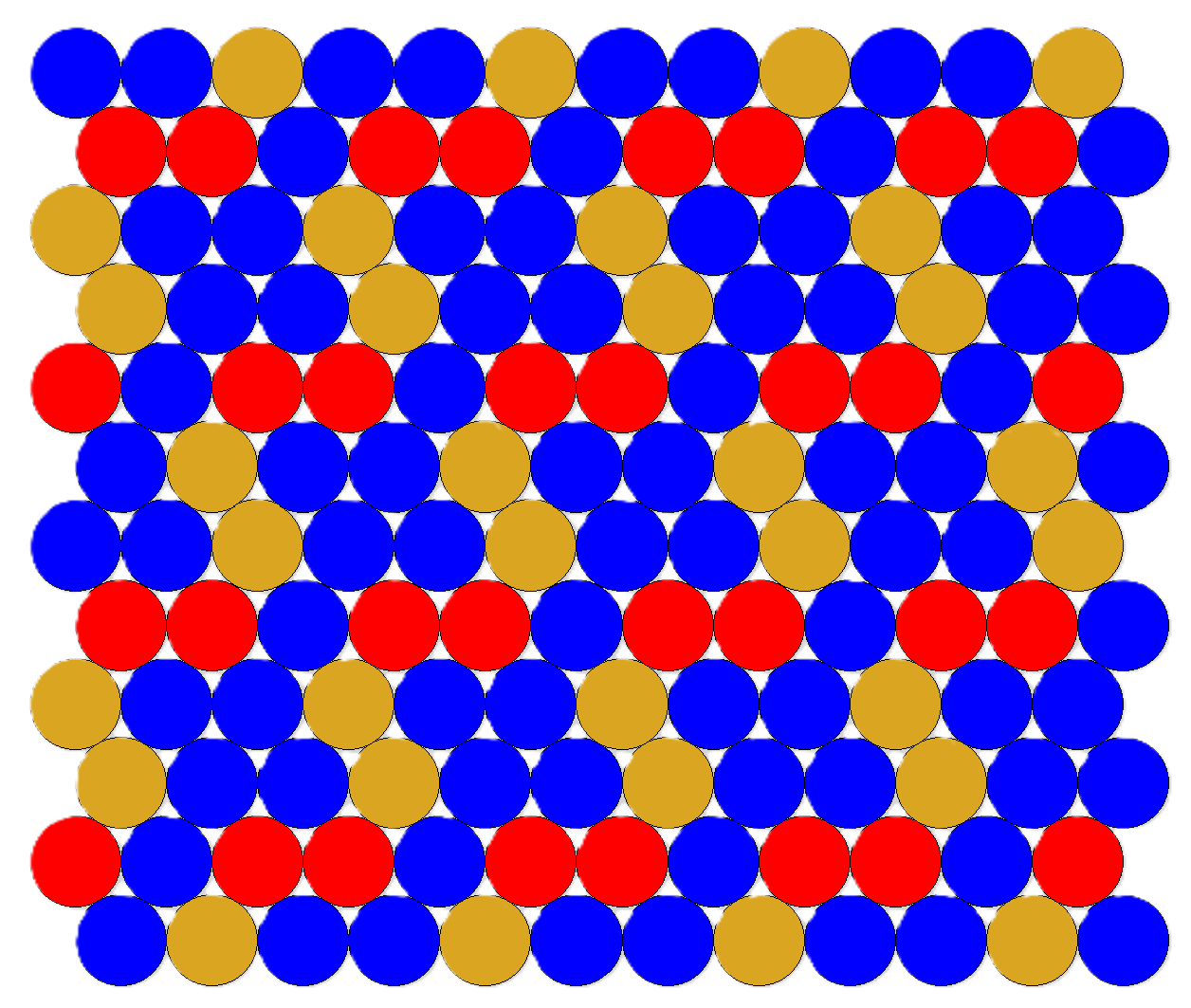}\\
    \centering \# 13 (24)
  \end{minipage}\hfill
  \begin{minipage}[t]{0.24\textwidth}
    \centering\includegraphics[height=3cm]{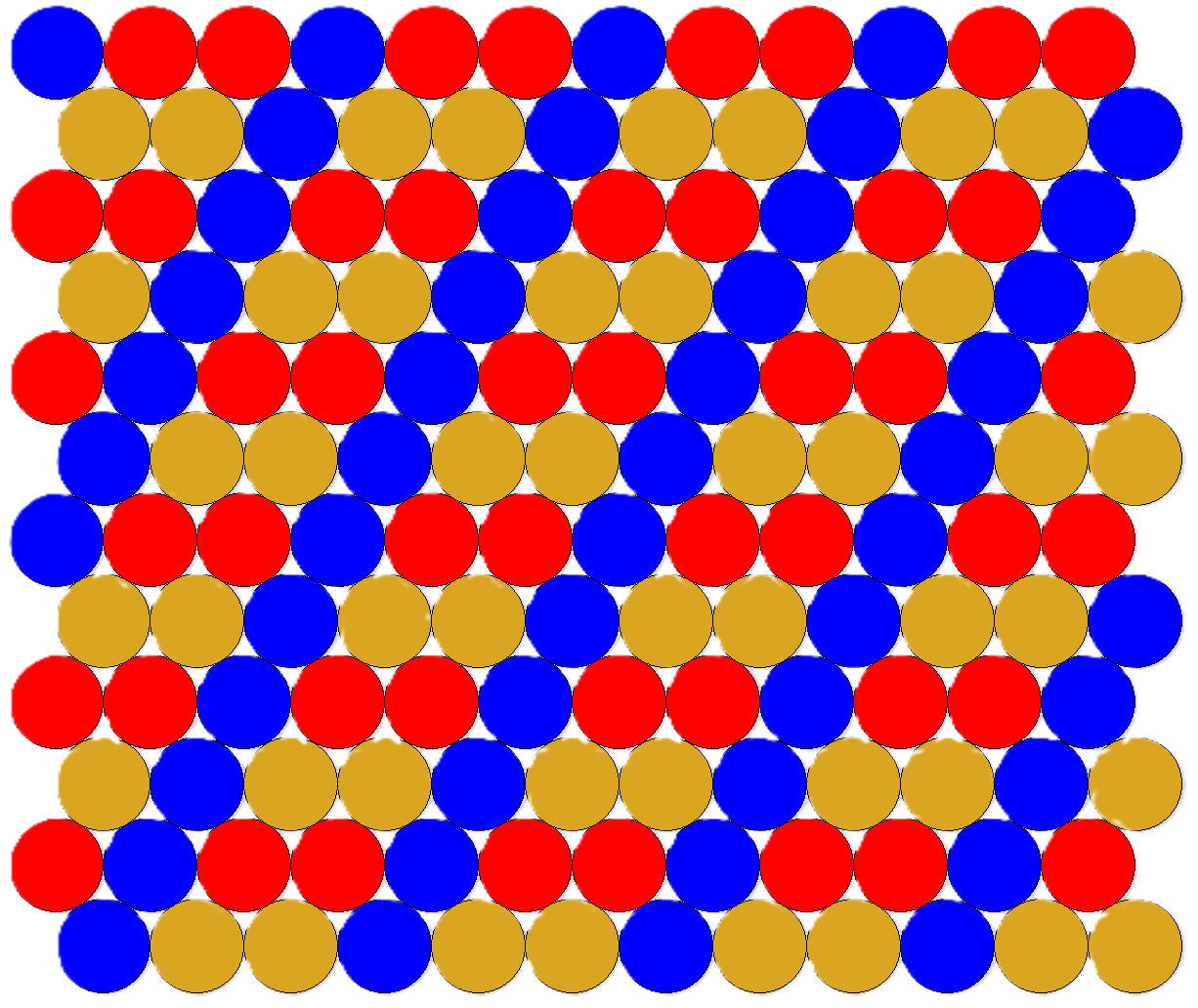}\\
    \centering \# 14 (26)
  \end{minipage}\hfill
  \begin{minipage}[t]{0.24\textwidth}
    \centering\includegraphics[height=3cm]{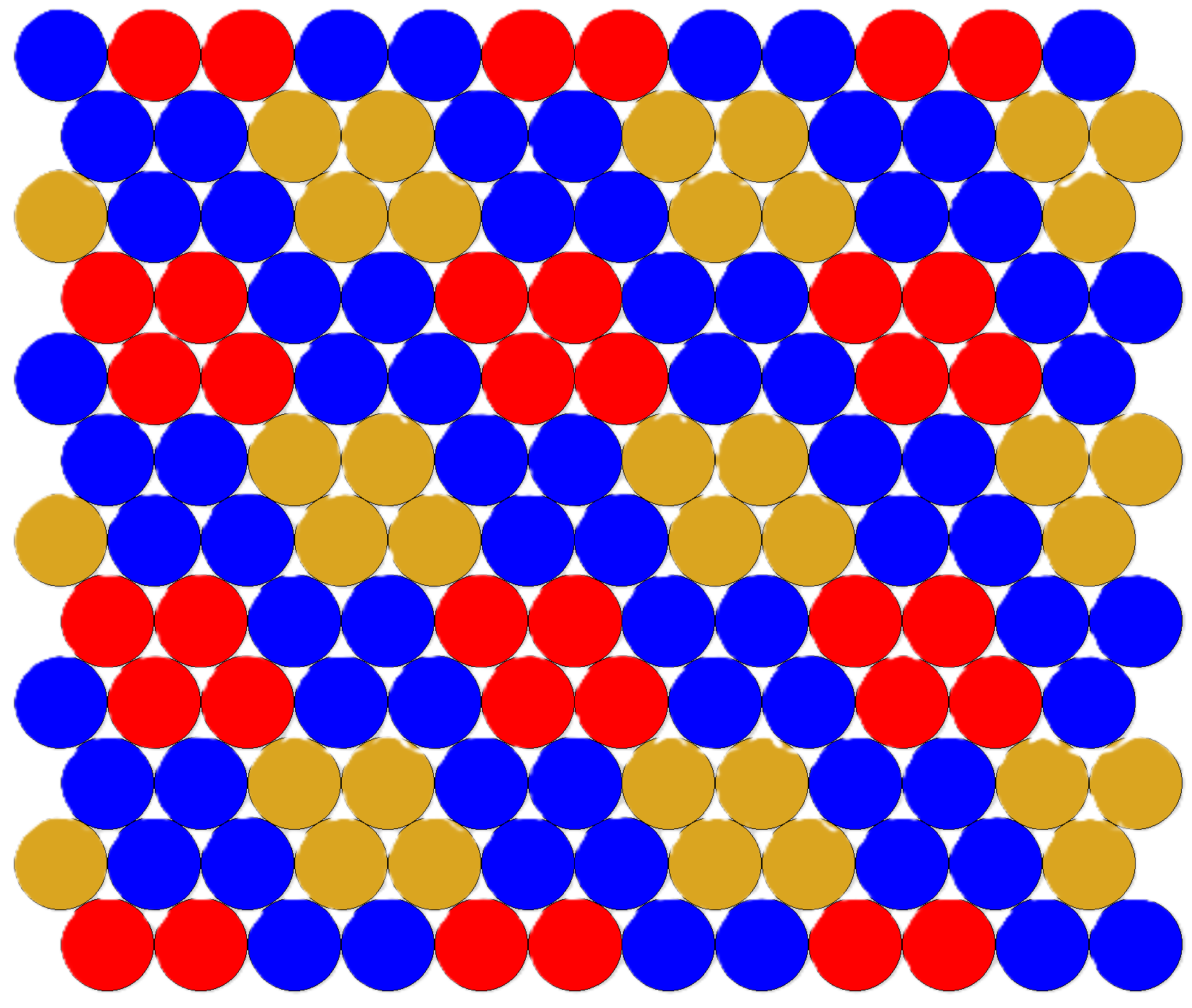}\\
    \centering \# 15 (35,38)
  \end{minipage}\hfill
  \begin{minipage}[t]{0.24\textwidth}
    \centering\includegraphics[height=3cm]{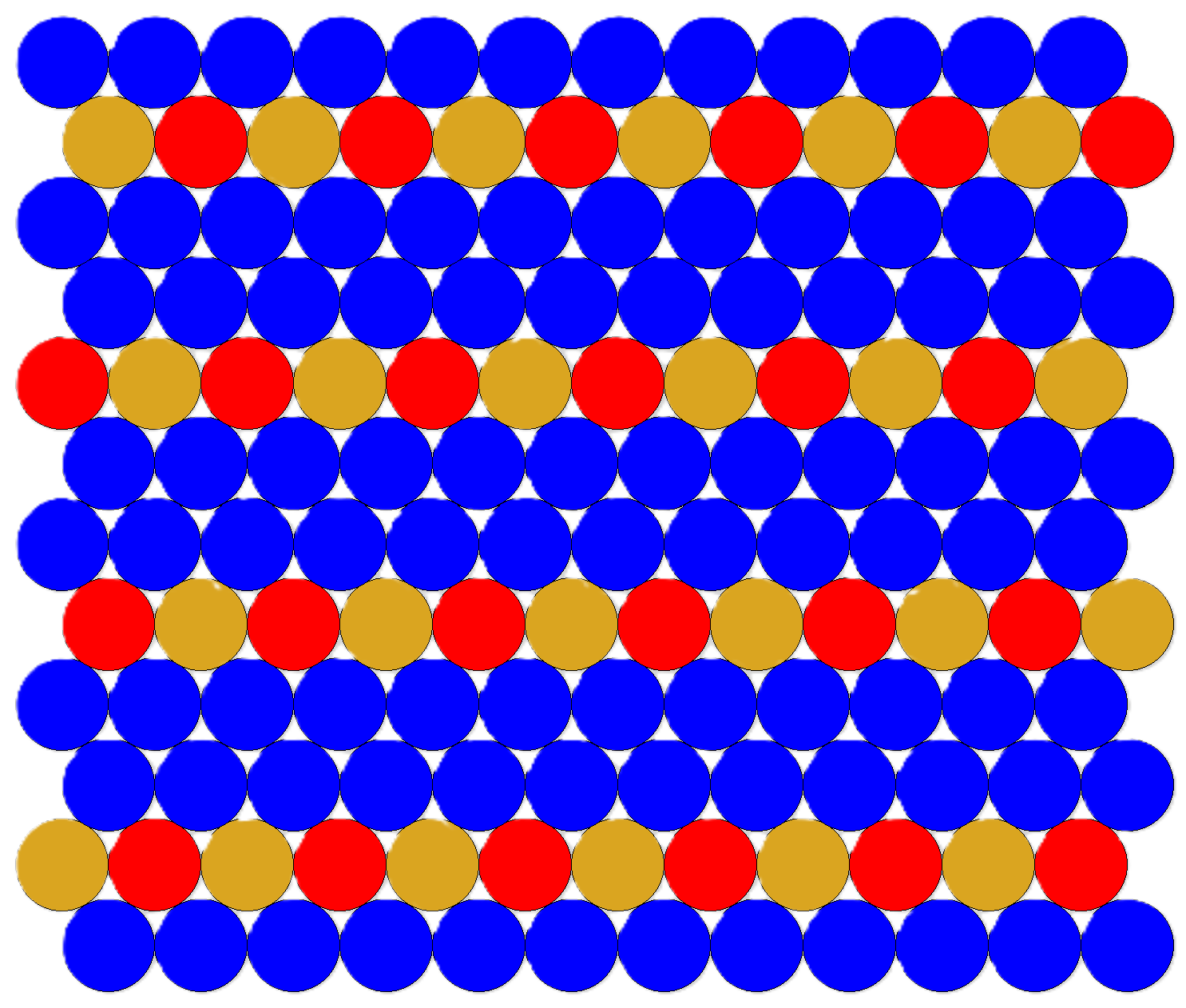}\\
    \centering \# 16 (26)
  \end{minipage}

  \caption{Catalog of regular stripes on the triangular grid (in this and the following figures, red and yellow symbols represent ``particles'' of species 1 and 2, respectively;
blue symbols: empty sites).
  Below each panel, we report within parentheses the respective case(s) in Table I.
  The panels have been organized according to the number of components present in a single stripe: one species (from \#1 to \#7) or two species (from \#8 to \#16).}
  \label{gallery1}
\end{figure*}

\begin{figure*}[htbp]
  \centering
  \begin{minipage}[t]{0.24\textwidth}
    \centering
    \includegraphics[height=2.5cm]{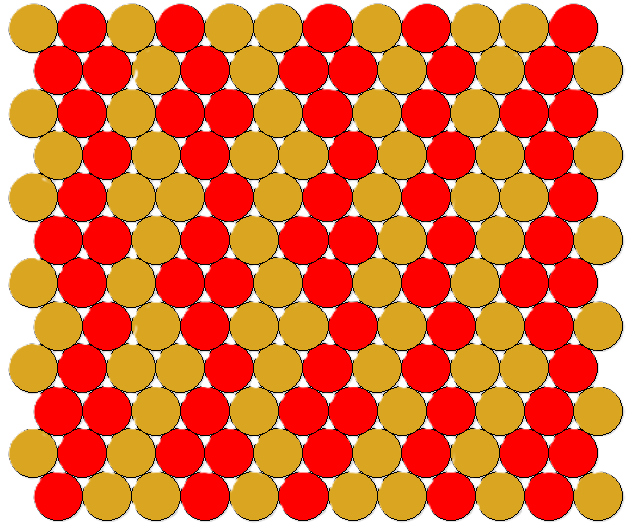}\\
    \# 1 (16,27)
  \end{minipage}\hfill
  \begin{minipage}[t]{0.24\textwidth}
    \centering
    \includegraphics[height=2.5cm]{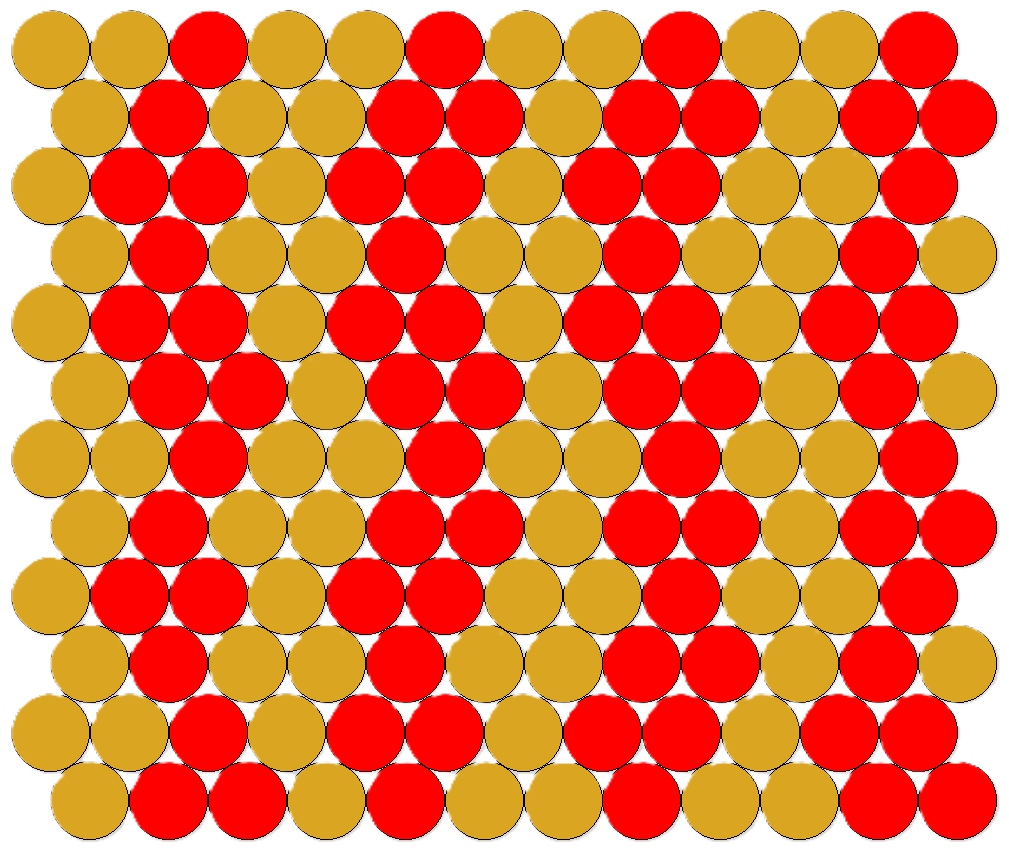}\\
    \# 2 (25,34)
  \end{minipage}\hfill
  \begin{minipage}[t]{0.24\textwidth}
    \centering
    \includegraphics[height=2.5cm]{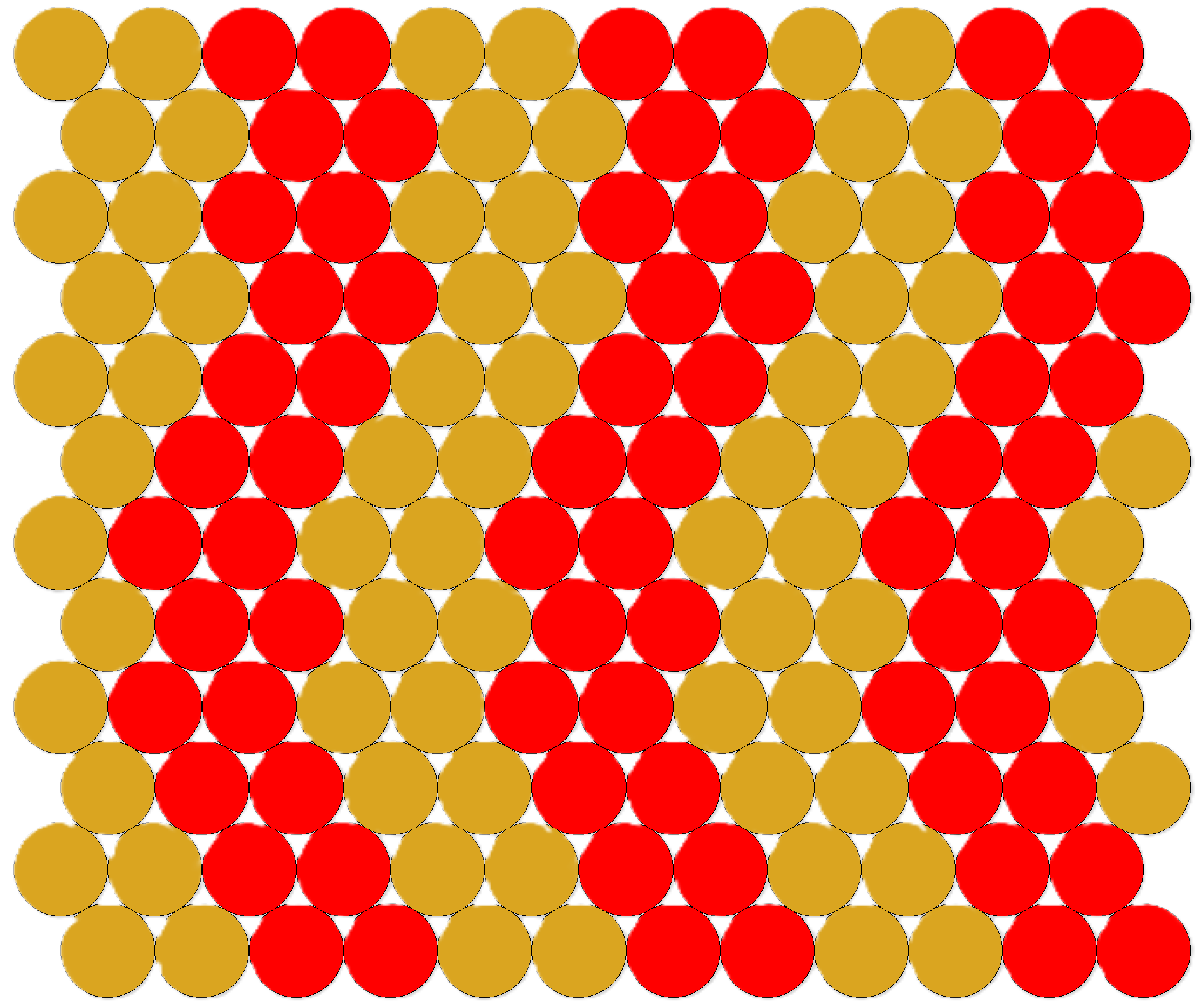}\\
    \# 3 (10,37,40)
  \end{minipage}\hfill
  \begin{minipage}[t]{0.24\textwidth}
    \centering
    \includegraphics[height=2.5cm]{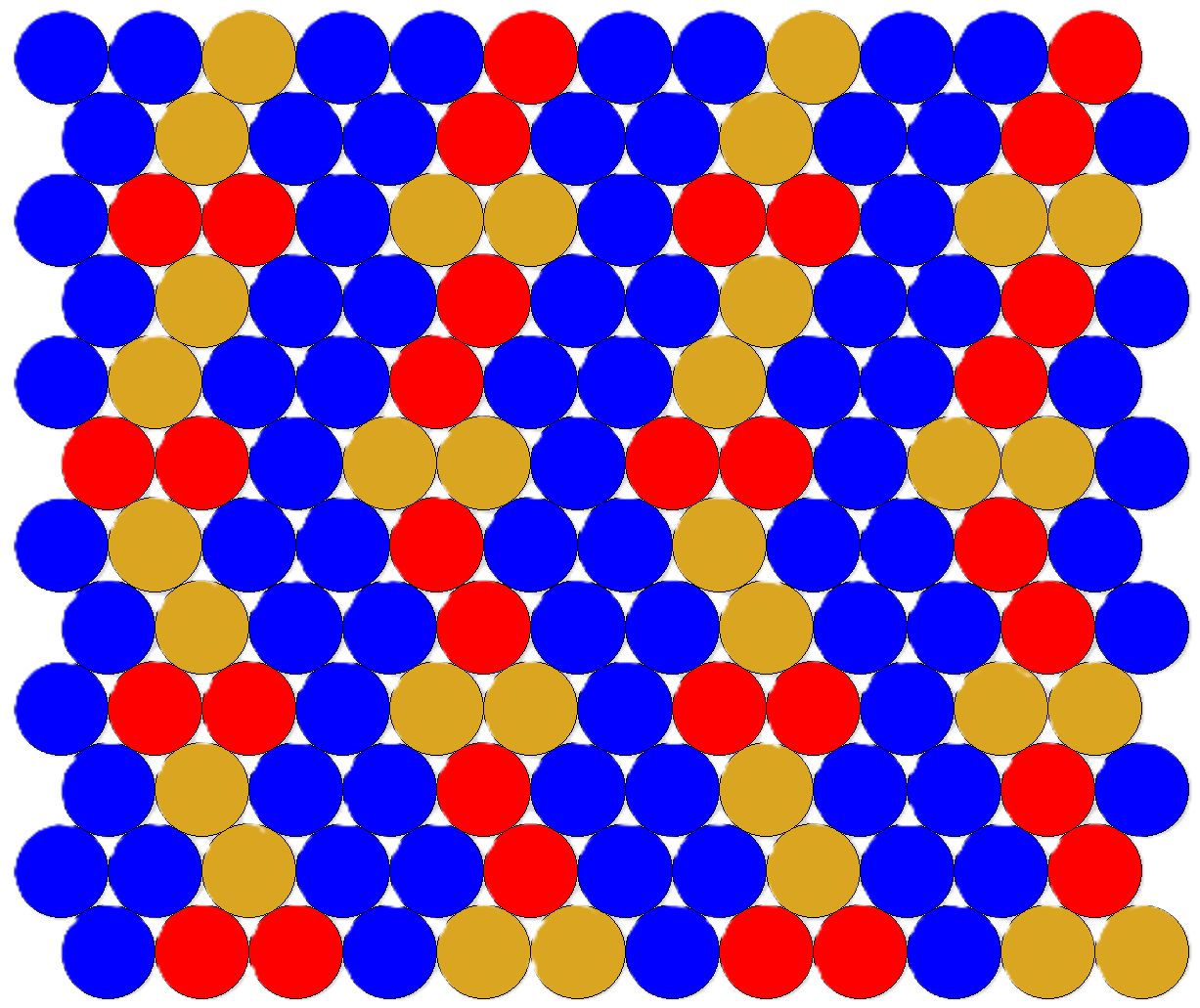}\\
    \# 4 (27)
  \end{minipage}
  \caption{Instances of irregular stripes.
Each of them is a particular state of a ``stripe liquid'' phase (see text).}
  \label{gallery2}
\end{figure*}

\begin{figure*}[htbp]
  \centering
  \begin{minipage}[t]{0.24\textwidth}
    \centering\includegraphics[height=2.5cm]{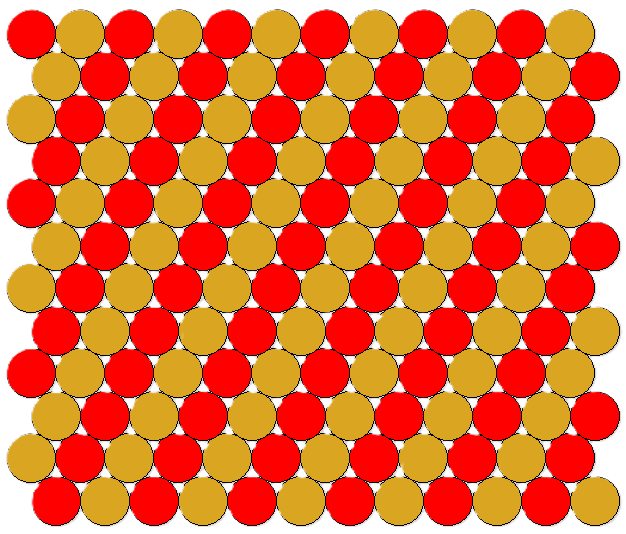}\\
    \# 1 (26)
  \end{minipage}\hfill
  \begin{minipage}[t]{0.24\textwidth}
    \centering\includegraphics[height=2.5cm]{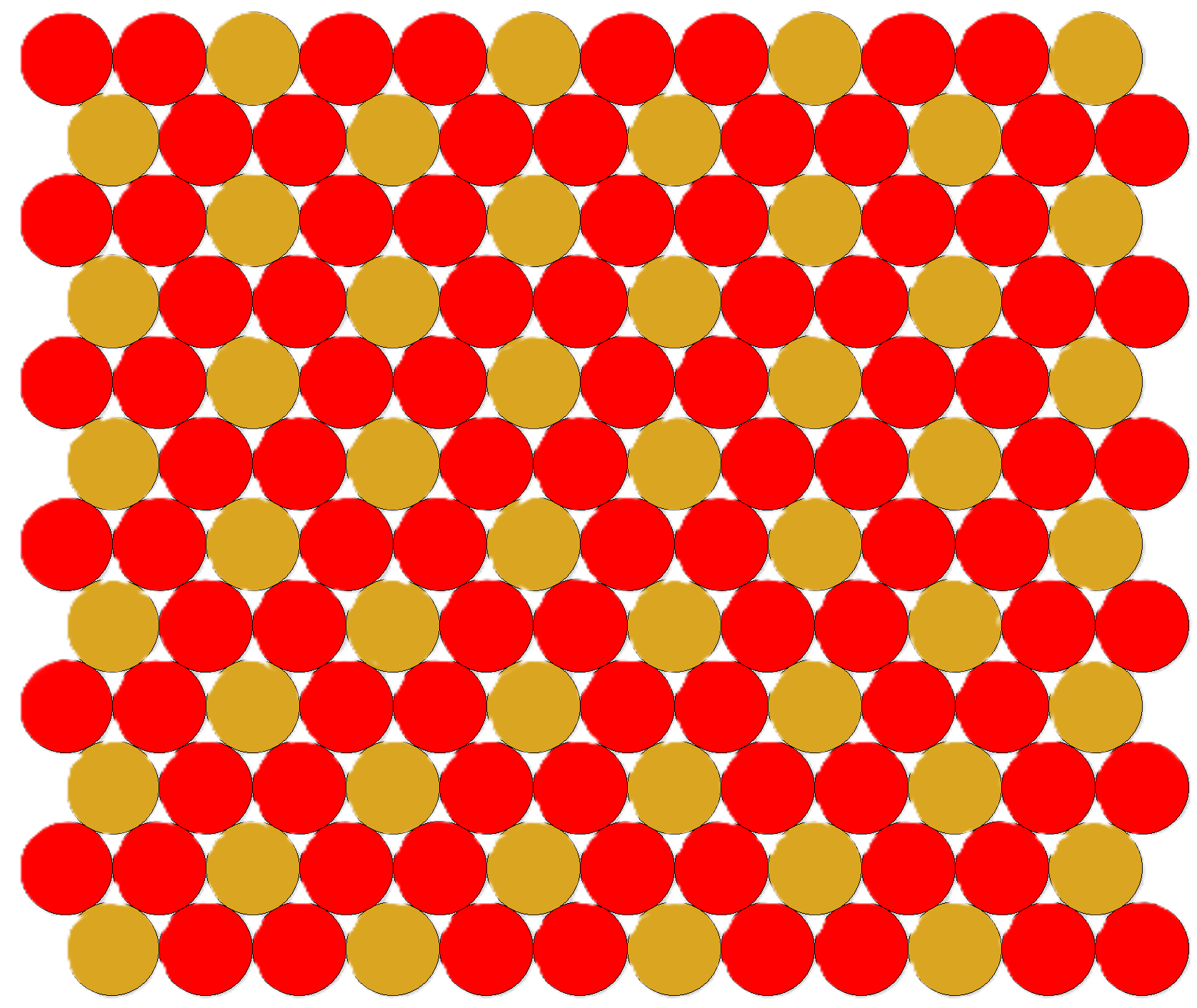}\\
    \# 2 (26)
  \end{minipage}\hfill
  \begin{minipage}[t]{0.24\textwidth}
    \centering\includegraphics[height=2.5cm]{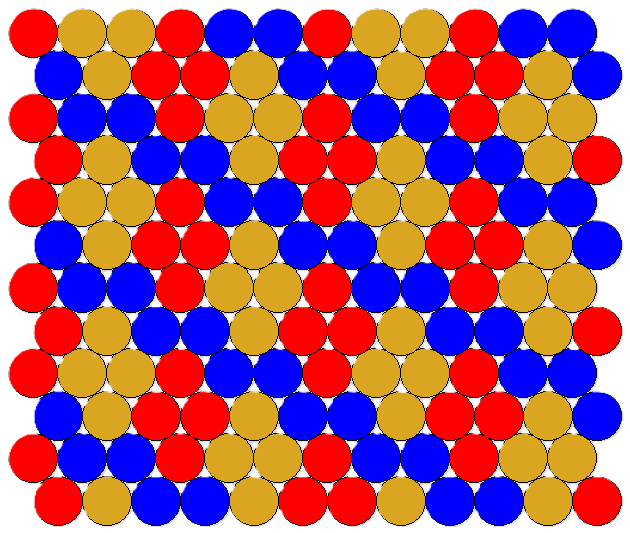}\\
    \# 1 (5)
  \end{minipage}\hfill
  \begin{minipage}[t]{0.24\textwidth}
    \centering\includegraphics[height=2.5cm]{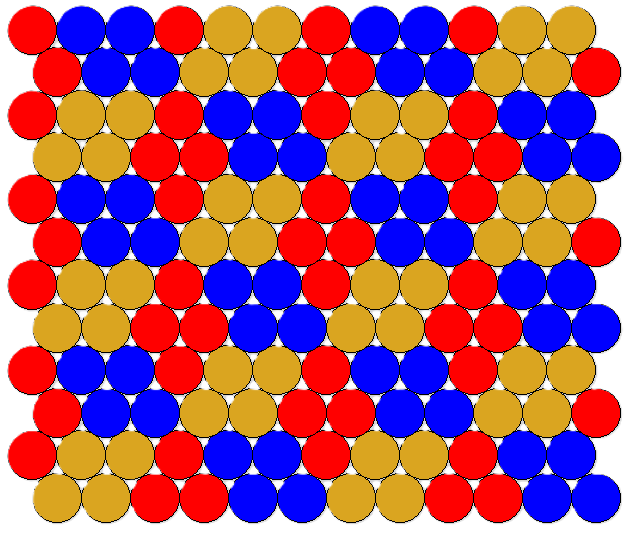}\\
    \# 2 (5)
  \end{minipage}\\ \vspace{1em}
  
  \begin{minipage}[t]{0.24\textwidth}
    \centering\includegraphics[height=2.5cm]{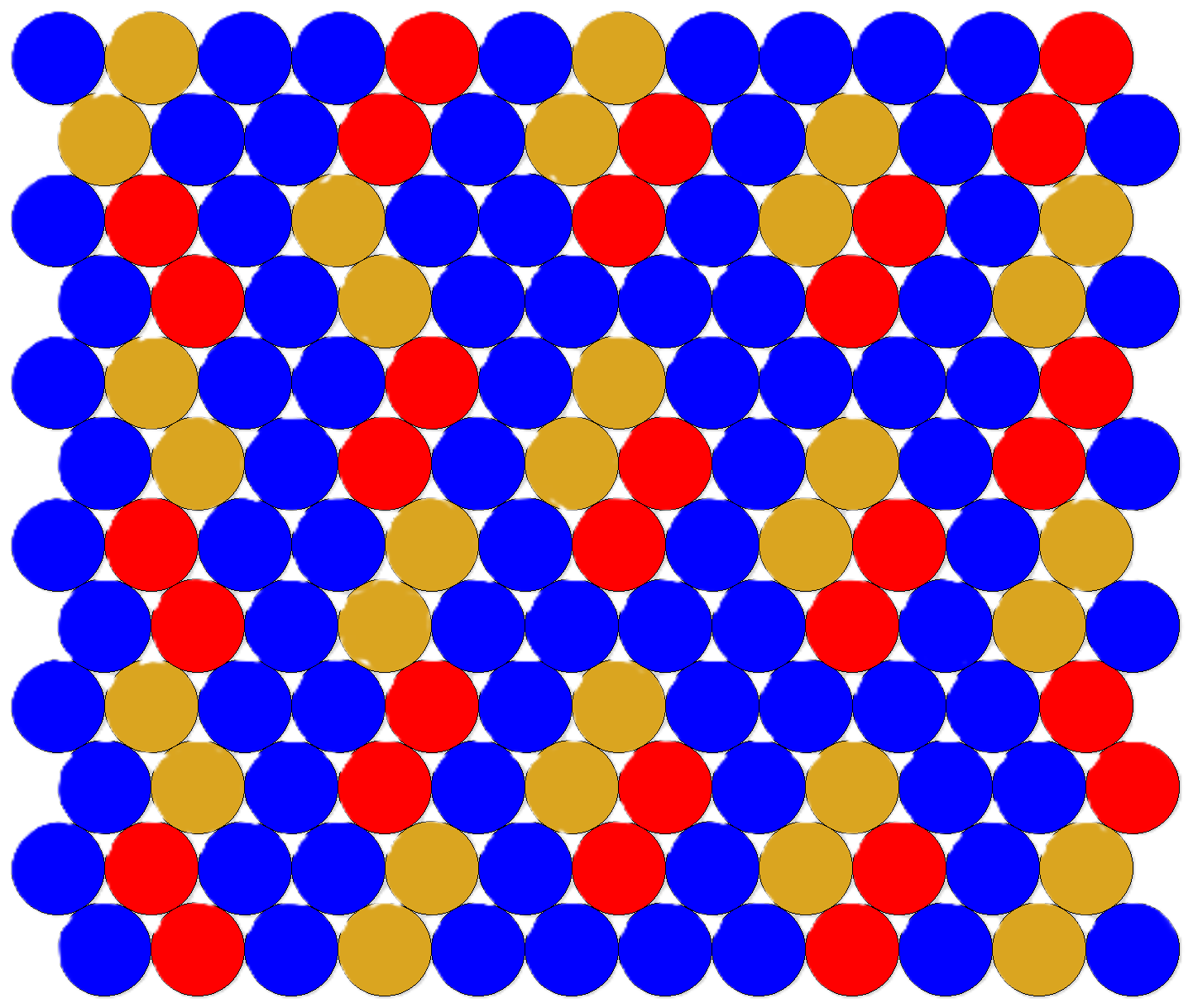}\\
    \# 1 (5)
  \end{minipage}\hfill
  \begin{minipage}[t]{0.24\textwidth}
    \centering\includegraphics[height=2.5cm]{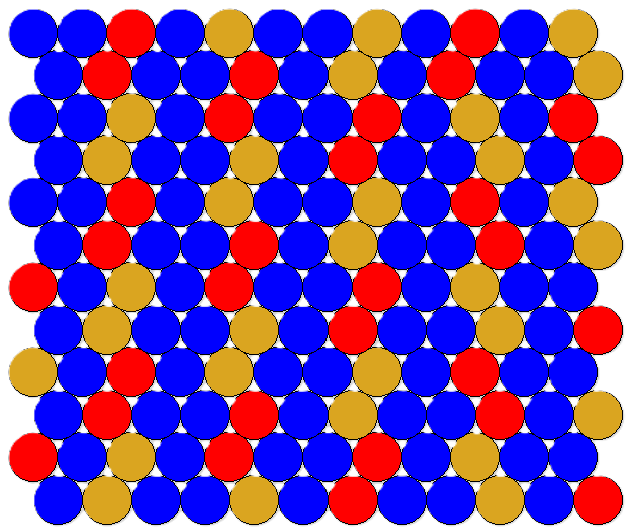}\\
    \# 2 (5)
  \end{minipage}\hfill
  \begin{minipage}[t]{0.24\textwidth}
    \centering\includegraphics[height=2.5cm]{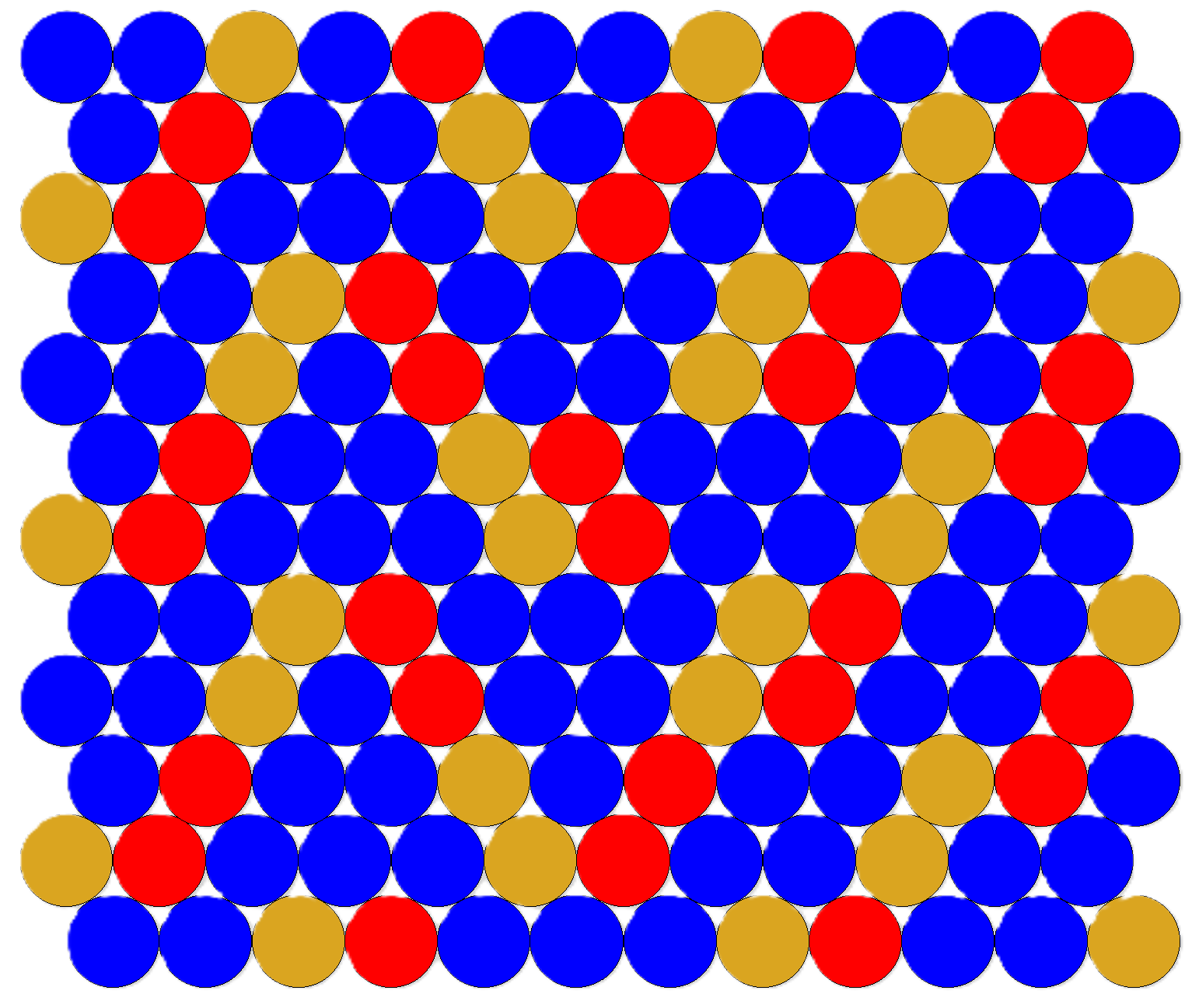}\\
    \# 3 (5)
  \end{minipage}\hfill
  \begin{minipage}[t]{0.24\textwidth}
    \centering\includegraphics[height=2.5cm]{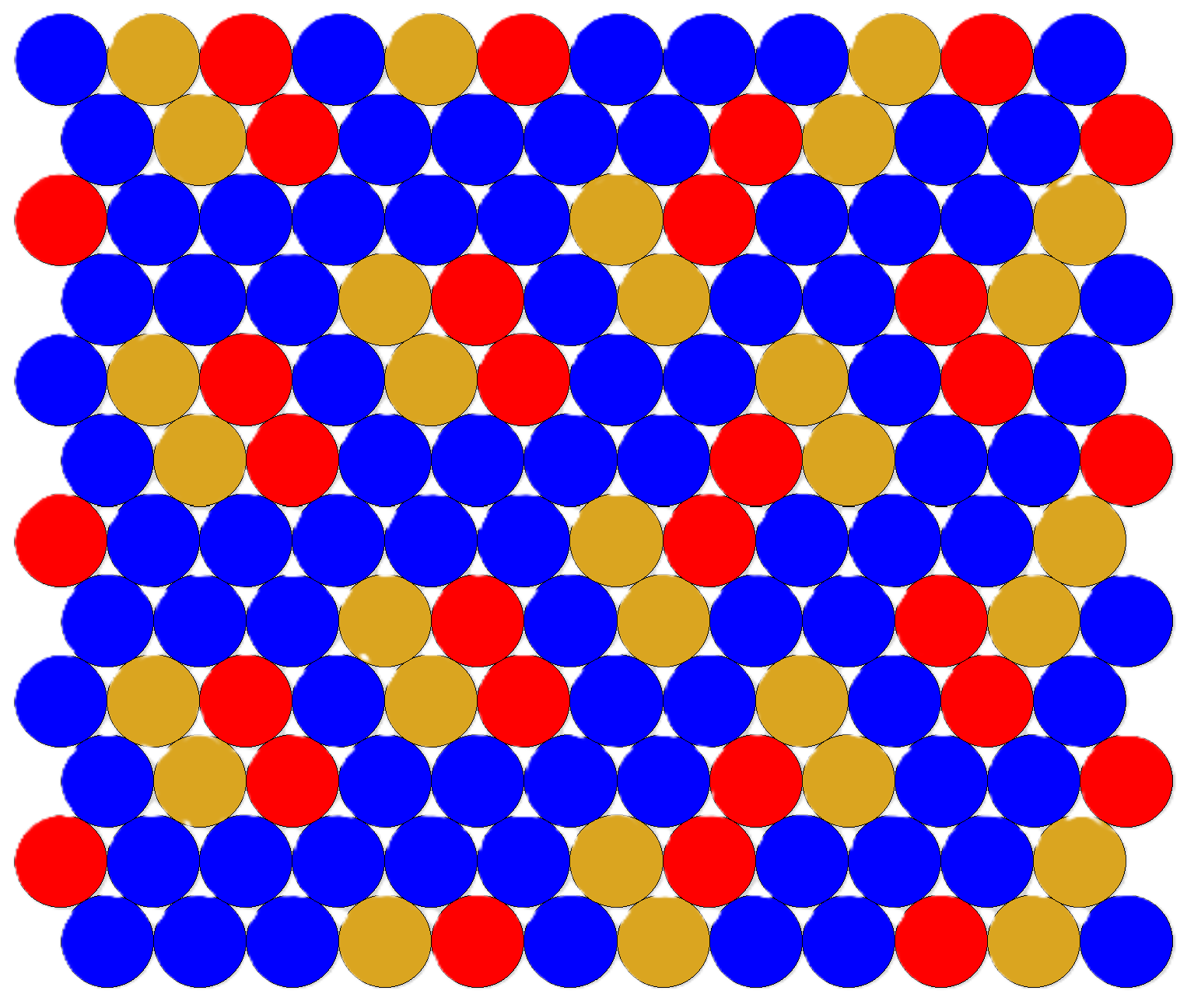}\\
    \# 4 (5)
  \end{minipage}\\ \vspace{1em}

  \begin{minipage}[t]{0.24\textwidth}
    \centering\includegraphics[height=2.5cm]{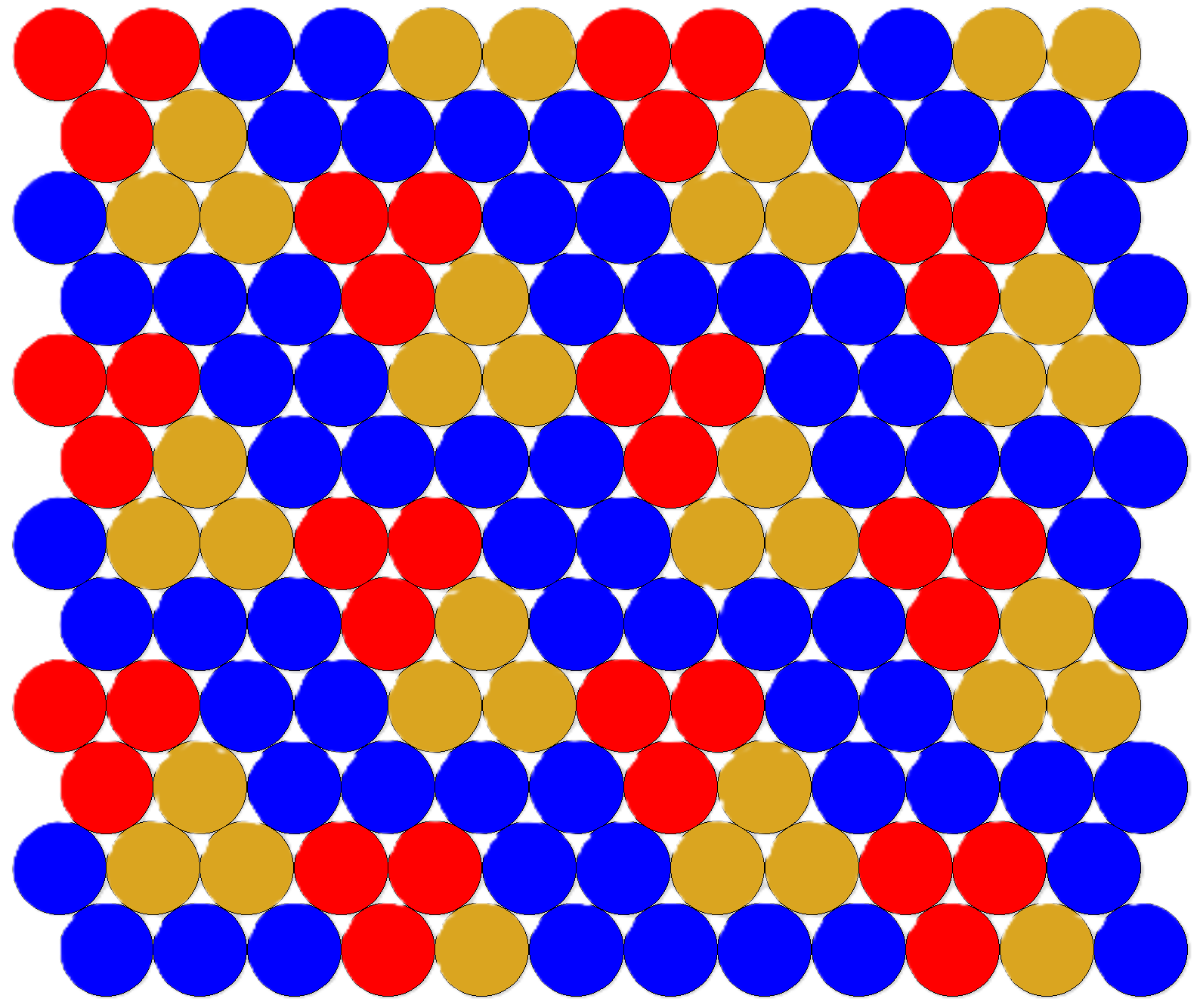}\\
    \# 1 (5)
  \end{minipage}\hfill
  \begin{minipage}[t]{0.24\textwidth}
    \centering\includegraphics[height=2.5cm]{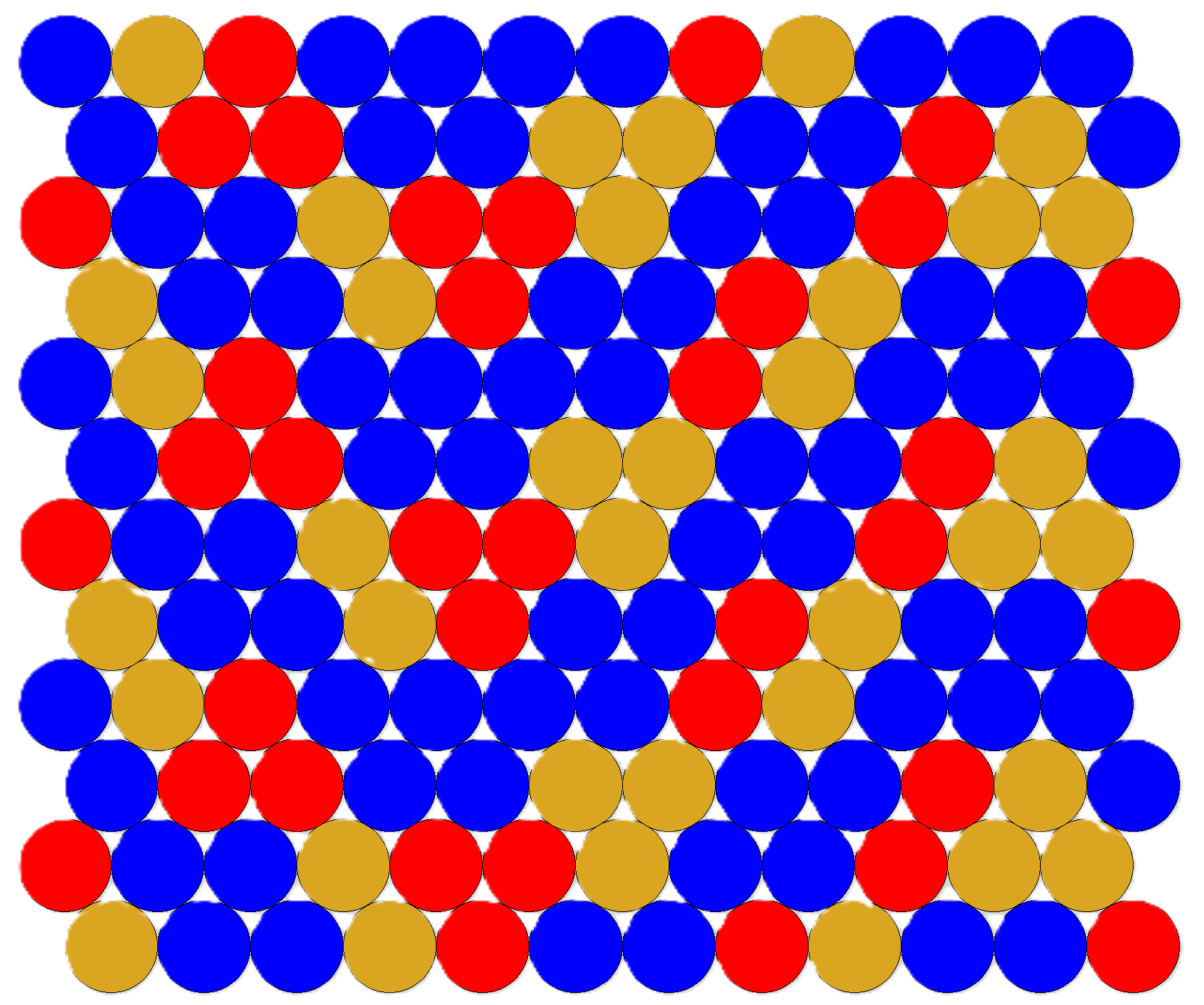}\\
    \# 2 (5)
  \end{minipage}\hfill
  \begin{minipage}[t]{0.24\textwidth}
    \centering\includegraphics[height=2.5cm]{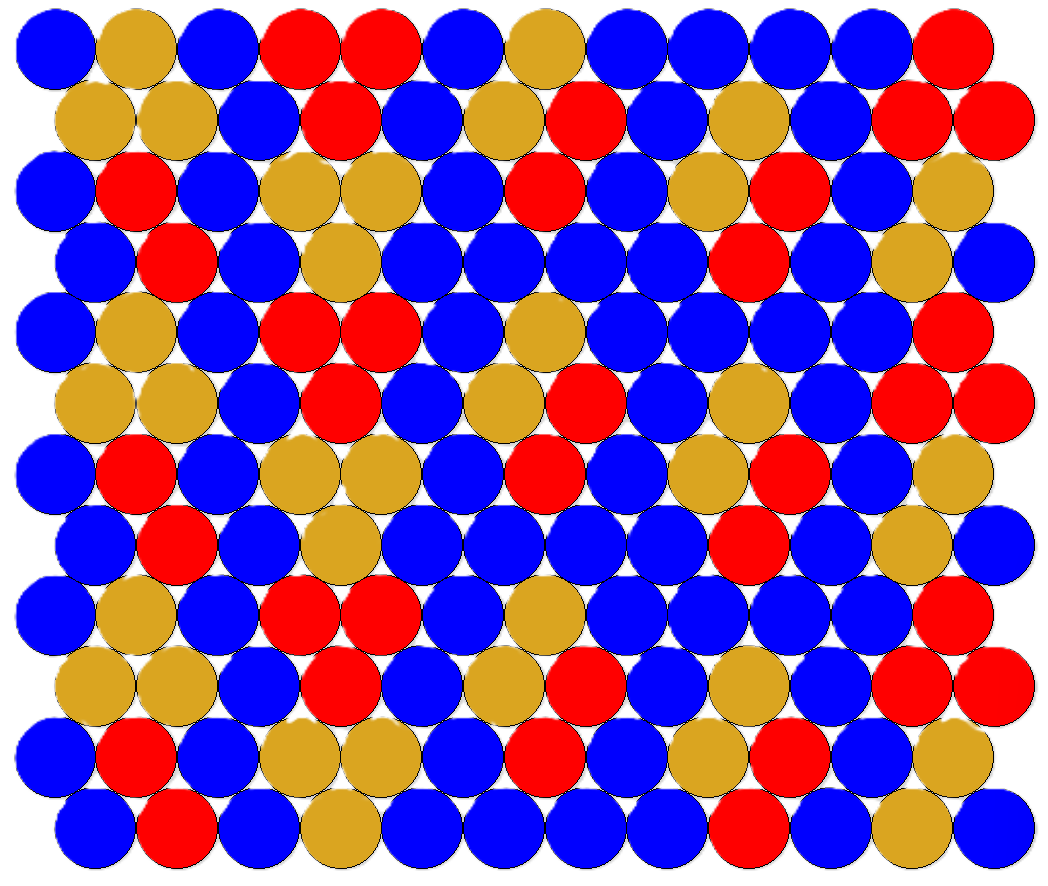}\\
    \# 3 (5)
  \end{minipage}\hfill
  \begin{minipage}[t]{0.24\textwidth}
    \centering\includegraphics[height=2.5cm]{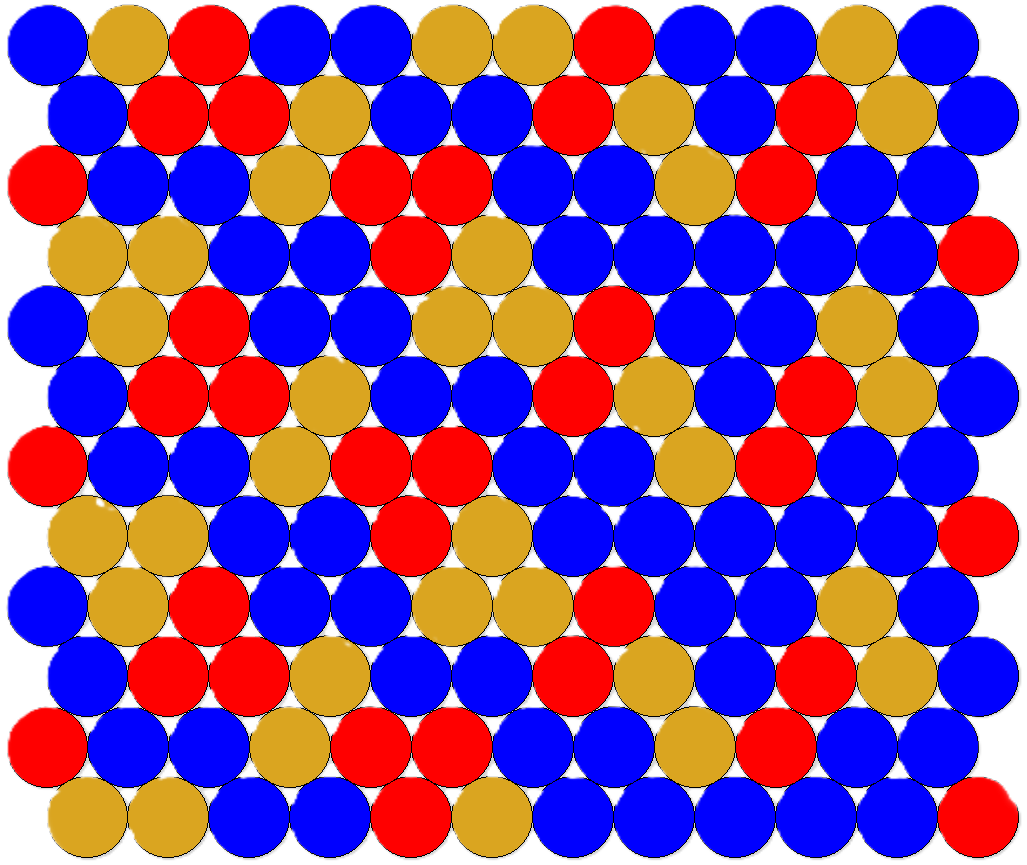}\\
    \# 4 (5)
  \end{minipage}\\ \vspace{1em}

  \begin{minipage}[t]{0.24\textwidth}
    \centering\includegraphics[height=2.5cm]{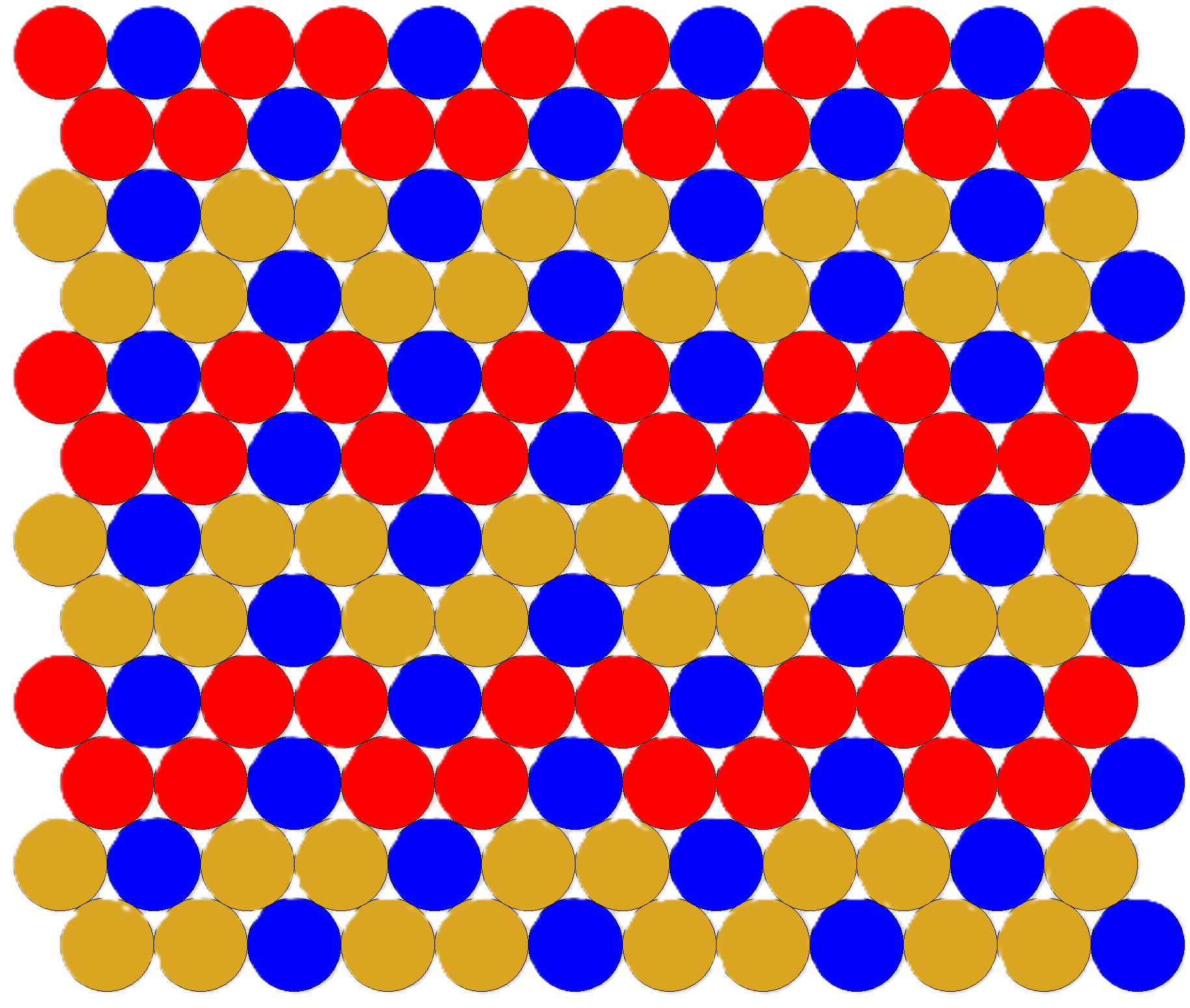}\\
    \# 1 (9)
  \end{minipage}\hfill
  \begin{minipage}[t]{0.24\textwidth}
    \centering\includegraphics[height=2.5cm]{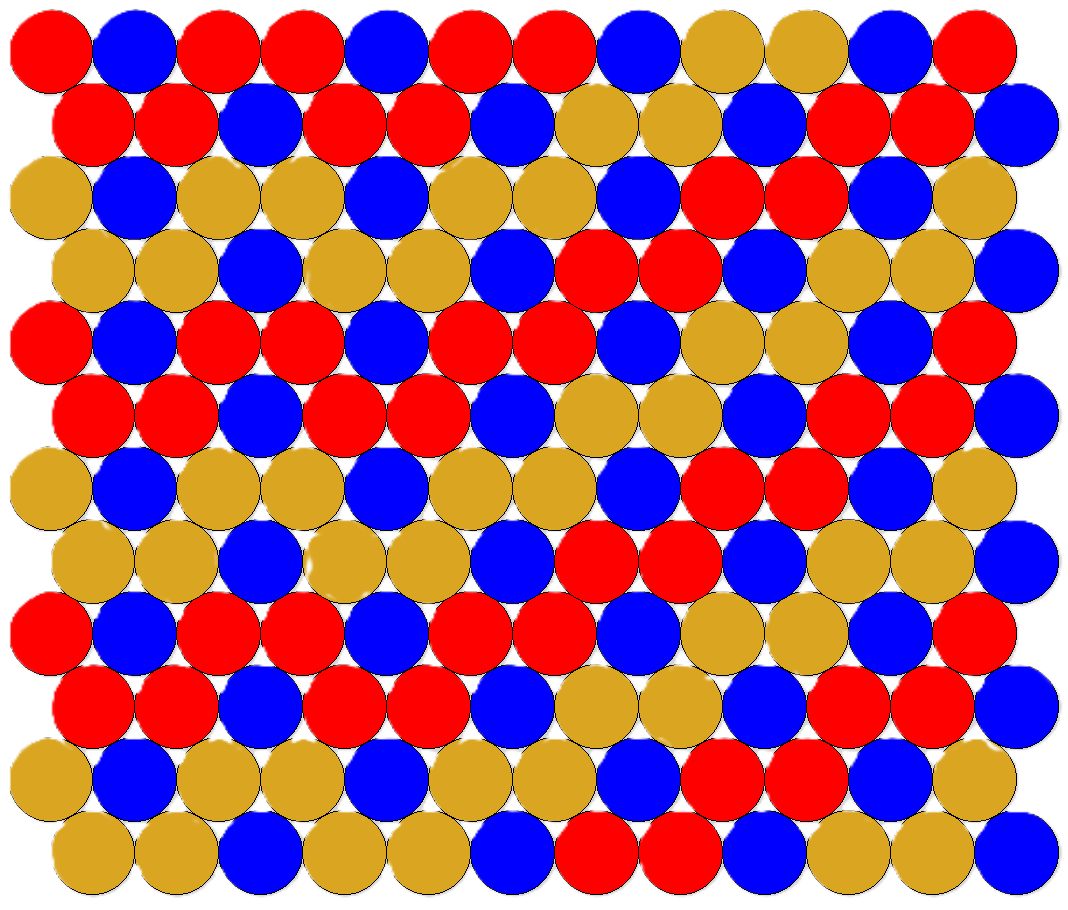}\\
    \# 2 (9)
  \end{minipage}\hfill
  \begin{minipage}[t]{0.24\textwidth}
    \centering\includegraphics[height=2.5cm]{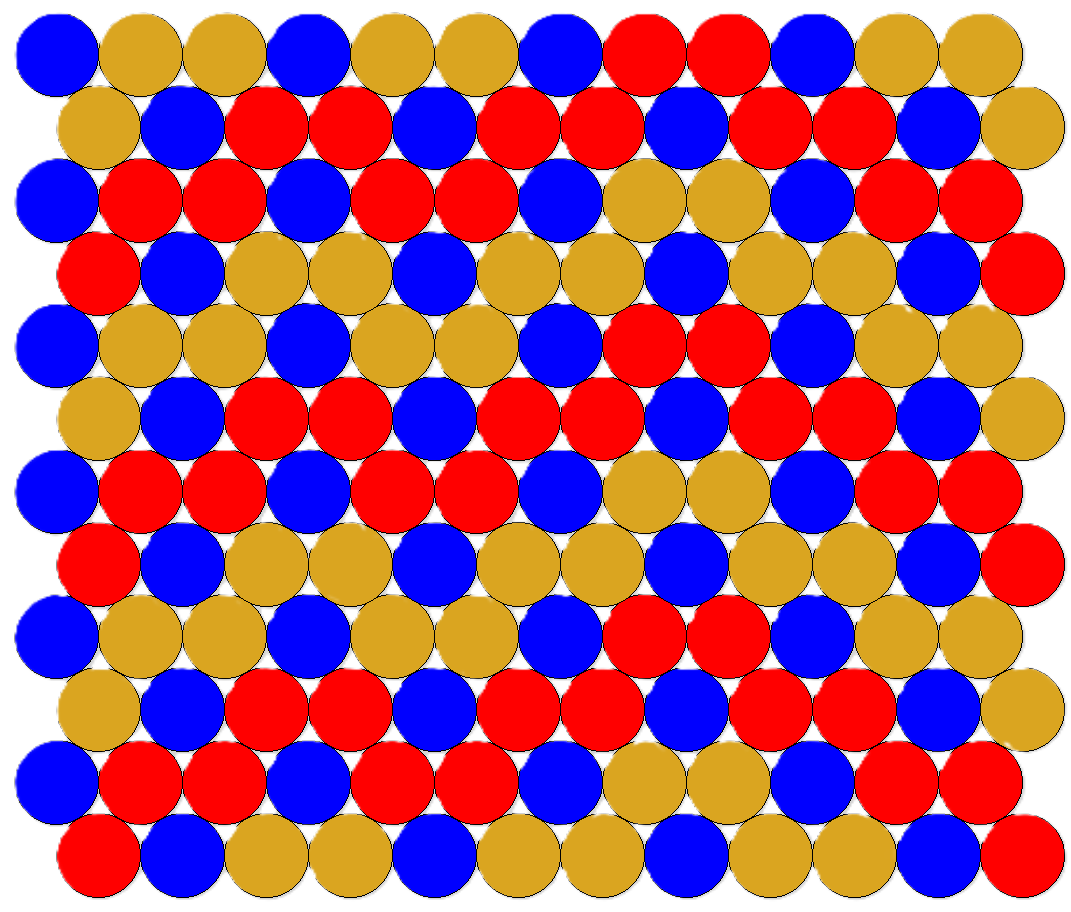}\\
    \# 3 (9)
  \end{minipage}\hfill
  \begin{minipage}[t]{0.24\textwidth}
    \centering\includegraphics[height=2.5cm]{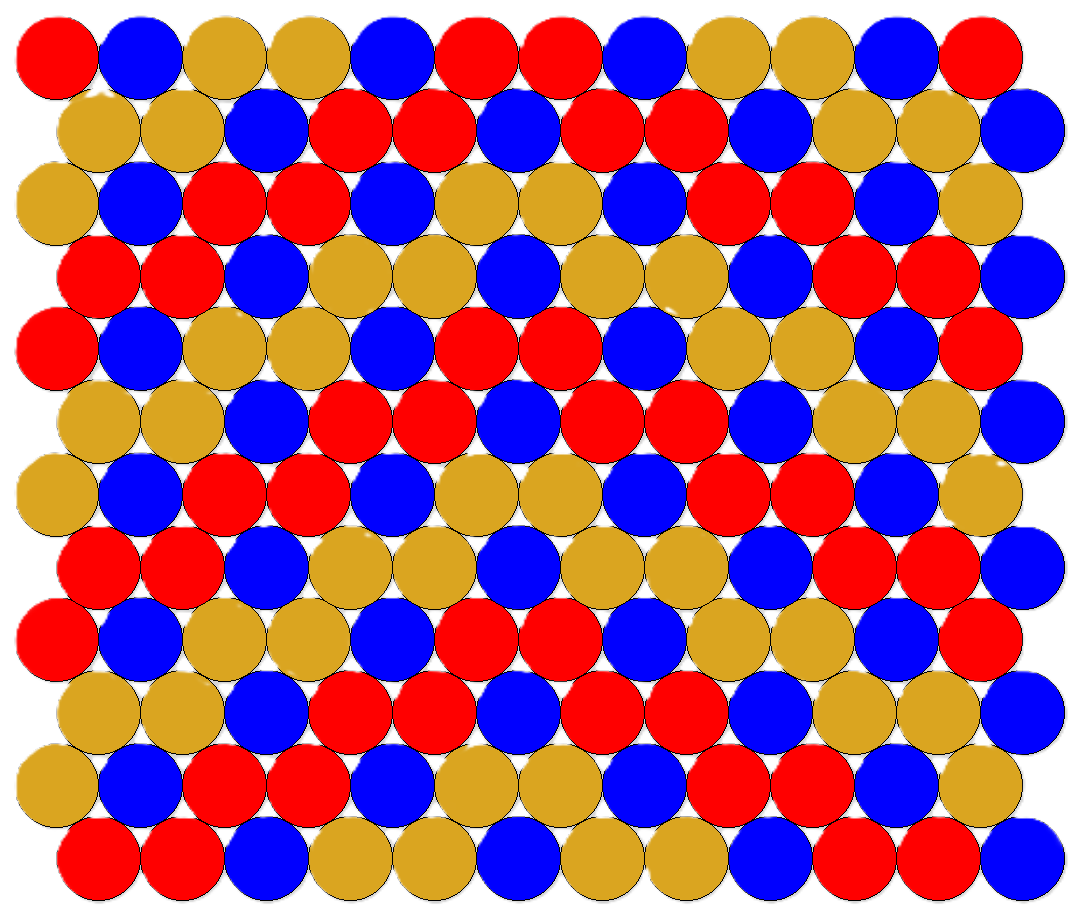}\\
    \# 4 (9)
  \end{minipage}
  \caption{This picture shows cases (within parentheses) where microstates of a different nature coexist (see text).
First row: ground states for case 26 (i.e., $U_{11}=(1,1,0)$ and $U_{12}=(-1,2,1)$, for $N=144,E=576$);
and for case 5 (i.e., $U_{11}=(1,1,1)$ and $U_{12}=(0,0,0)$, for $N=96,E=168$).
Rows 2 and 3: ground states for case 5, for $N=60,E=36$ and $N=72,E=72$, respectively.
Bottom row: ground states for case 9 (i.e., $U_{11}=(0,0,0)$ and $U_{12}=(1,-1,0)$, for $N=96,E=-144$).}
  \label{gallery3}
\end{figure*}

\section{Results: Symmetric mixtures}

In this Section we provide an overview of the patterns emerging in symmetric lattice-gas mixtures with $\mu_1=\mu_2$ when combining, in many possible ways, like and unlike interactions of the three aforementioned categories (obviously, the considered instances are necessarily limited in number).
Our results are outlined in Table I for the mixture defined on the triangular grid;
the patterns described in words in Table I are illustrated graphically in the Supplementary Material and there compared with the self-assembly structures observed, under the same conditions, on the spherical grid.
We will make evident that most structures on the triangular grid look locally similar on the spherical grid, but the latter grid remains unique in allowing the formation of patterns exhibiting rotation symmetries that could fruitfully be exploited for nanomaterials technology.
In the following, our attention will be focused on stripe patterns, while other kinds of self-assembly structures are touched upon only briefly at the end of the section.
We start considering two cases with $u^{(n)}_{1,1}=0$ and non-negative $u^{(n)}_{1,2}$ (cases 1 and 2 in Table I).
With two species unwilling to mix together, the low-density aggregates are irregular clusters of either ``color''.
At larger densities, the only way to keep the energy lowest (i.e., zero) is by expelling one species from the grid.
Cases 3-5 are also special somehow, and thus discussed separately from the other.
Here, like particles are mutually repelling, while no (off-core) interaction exists between unlike particles.
As the density grows at low $T$, keeping particles of same species spatially separate from each other becomes increasingly difficult, until, above a certain density, the energy is minimized by forming stripes, at least provided that like repulsion extends beyond nearest neighbors.
Stripes are preferred to other arrangements because they ensure the least content of total repulsive energy.
This concept is easily grasped in case 4 by looking at the completely filled grid ($N=144$):
sitting on an arbitrary particle, the number of first- or second-neighbor particles of same species is two in case of one-row stripes, while being three on average when the environment is compositionally disordered.
However, this is not enough in case 3, where many distinct linear arrangements of like particles have the same energy of stripes (see Supplementary Material).
On the other hand, in case 5 several kinds of stripes are possible.
In particular, the stripes for $N=144$ are thicker than in case 4, while two different stripe patterns coexist for $N=96$.
Clearly, this rich self-assembly behavior could only hardly have been anticipated from the shape of interaction, whereas it comes about automatically from the MCWL analysis.
The interactions giving rise to stable stripe order in lattice-gas mixtures are actually many:
A non-exhaustive list of possibilities can be found in Table I (for the triangular grid) and in the Supplementary Material.
By scrolling through this file, one immediately realizes that stripes occur in many shapes and, moreover, that the same typology of stripes can arise from several interactions.
We have prepared a few pictures to help the reader follow our discussion, by assembling graphical material taken from the Supplementary Material.
In particular, Fig.~\ref{gallery1} shows a gallery of {\em regular stripe patterns} on the triangular grid, together with the indication of the interaction(s) having these configurations as unique ground states (up to symmetry operations) in a $\mu$ interval;
any such state will thus correspond to a stripe phase in the thermodynamic limit.
We clarify that the list of patterns in Fig.~\ref{gallery1} is necessarily incomplete, since stripes incommensurate with a $12\times 12$ grid cannot appear.
For each pattern in Fig.~\ref{gallery1}, we can recognize a similar texture of the spherical grid that is inevitably less regular (these patterns are shown in the Supplementary Material).
In a separate figure (Fig.~\ref{gallery2}), we show instances of {\em irregular planar stripes} that are intrinsically degenerate at zero temperature, since being characterized by many similar, not symmetry-related, disordered microstates.
While the cases shown in Fig.~\ref{gallery1} would be representative of stripe ``solids'', each pattern in Fig.~\ref{gallery2} is paradigmatic of a {\em stripe liquid}, i.e., a stripe phase characterized by residual entropy at $T=0$.
Finally, in Fig.~\ref{gallery3} we see configurations relative to ``phases'' with zero-point entropy, due to the joint presence of patterns of a different nature.
With Figs.~\ref{gallery1}-\ref{gallery3} we have still been limited to a taxonomy of stripes, without entering in the motivations behind the stabilization of the various structures.
Indeed, this is a daunting task;
what will be attempted below is to identify --- in selected cases --- the interaction features favorable to the occurrence of stripes.
Far more complicate, if not even impossible, would be to prove analytically, in each specific case where stripes are expected to occur at low temperature, that all alternative structures are indeed far from optimal.
Let us first analyze the regular stripe patterns in Fig.~2.
As a first remark, we note that these patterns can be classified according to the number of components making a single layer.
In mixtures of distinct molecular species, the periodic alternation of one-color and two-color layers would indeed correspond to qualitatively different phases.
However, also in symmetric mixtures a distinction can be made between the two types of stripe phases, by saying that only in a two-color layer the symmetry between the species is not broken.
Take, for instance, the case of stripes \# 1 (in the classification proposed in Fig.~\ref{gallery1}).
We have already discussed the origin of these stripes in case 4.
Since the energy per particle is $u^{(1)}_{1,1}+u^{(2)}_{1,1}+3u^{(3)}_{1,1}+2(u^{(1)}_{1,2}+u^{(2)}_{1,2})$, stripes \# 1 are also favored when a short-range like repulsion is more than compensated by a longer-range like attraction and/or a short-range unlike attraction.
In detail, stripes \# 1 are present when the like repulsion is extended to second neighbors and the unlike interaction is zero or attractive (cases 4, 22, 23).
We have also cases where the like repulsion is zero o short range, e.g., $U_{11}=(1,0,0)$, but the unlike attraction should then be extended to at least second neighbors (cases 7, 8, and 15).
In particular, cases 7 and 8 concern two distinct species of hard-core particles interacting through a SW attraction of relatively long range.
In these cases, the phase diagram features a single, first-order-like transition between vapor and stripe solid~\cite{prestipino2023density}.
The thickness of stripes grows with the range of the attraction.
In the intermediate ``coexistence region'', the shape of the solid droplet changes upon compression according to the same universal pathway identified for one-component fluids~\cite{Binder2012,Prestipino2015,Abramo2015}.
If the unlike attraction is purely first-neighbor, stripes only occur in the presence of an unlike repulsion between third neighbors (cases 13 and 21) --- actually irrelevant for stripes, but harmful to competing configurations where pairs of unlike third neighbors are present.
For example, stripes \# 1 are stable for $U_{11}=(1,0,0)$ and $U_{12}=(-1,0,1)$, but not for $U_{11}=(1,0,0)$ and $U_{12}=(-1,0,0)$;
similarly, the mixture with $U_{11}=(0,0,0)$ and $U_{12}=(-1,0,1)$ has stable stripes, in contrast to the one with $U_{11}=(0,0,0)$ and $U_{12}=(-1,0,0)$.
Furthermore, if we take $U_{11}=(1,0,0)$ and $U_{12}=(-1,0,0)$, where stripes are not present, and add a like attraction between third neighbors, then the interaction becomes stripe-forming (case 33).
Another enlightening example is $U_{11}=(1,1,0)$ and $U_{12}=(-1,0,1)$, which is so inclined to making stripes that the latter are also promoted when we include a strong repulsion between unlike second neighbors --- as in case 26, where $U_{11}=(1,1,0)$ and $U_{12}=(-1,2,1)$, although here stripes are degenerate with a crystalline state.
When stripes \# 2 are considered instead, the energy per particle is $2u_{1,1}^{(1)}+u_{1,1}^{(2)}+u_{1,1}^{(3)}+u_{1,2}^{(1)}+2(u_{1,2}^{(2)}+u_{1,2}^{(3)})$, meaning that these stripes are particularly favored either by a short-range like attraction or by a longer-range unlike attraction, as indeed observed in cases 9, 34,35, 38, and 39.

\begin{figure*}[htbp]
  \centering
  \begin{minipage}[t]{0.24\textwidth}
    \centering\includegraphics[height=3cm]{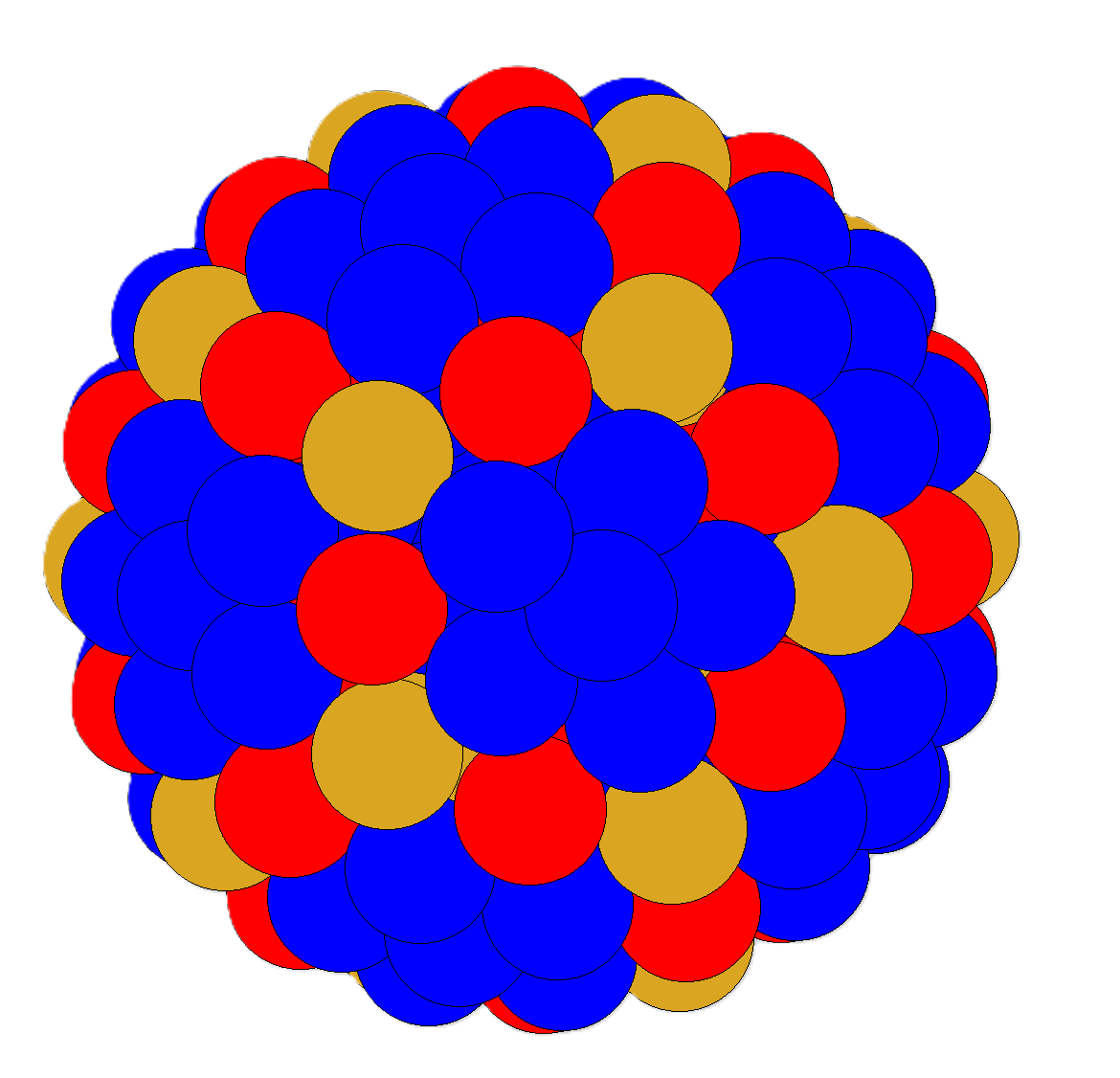}\\
    \# 1 (19,21)
  \end{minipage}\hfill
  \begin{minipage}[t]{0.24\textwidth}
    \centering\includegraphics[height=3cm]{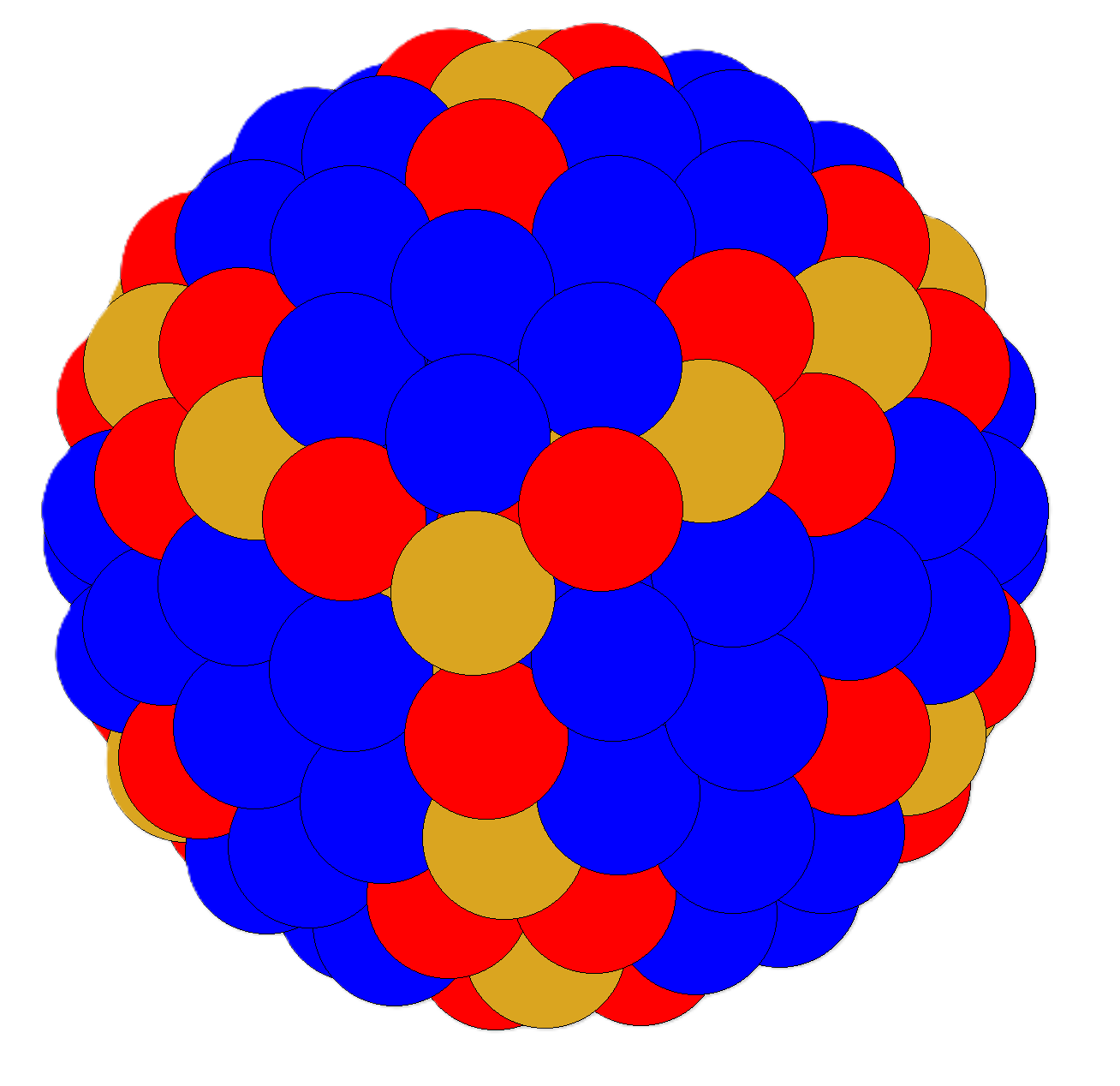}\\
    \# 2 (11,19)
  \end{minipage}\hfill
  \begin{minipage}[t]{0.24\textwidth}
    \centering\includegraphics[height=3cm]{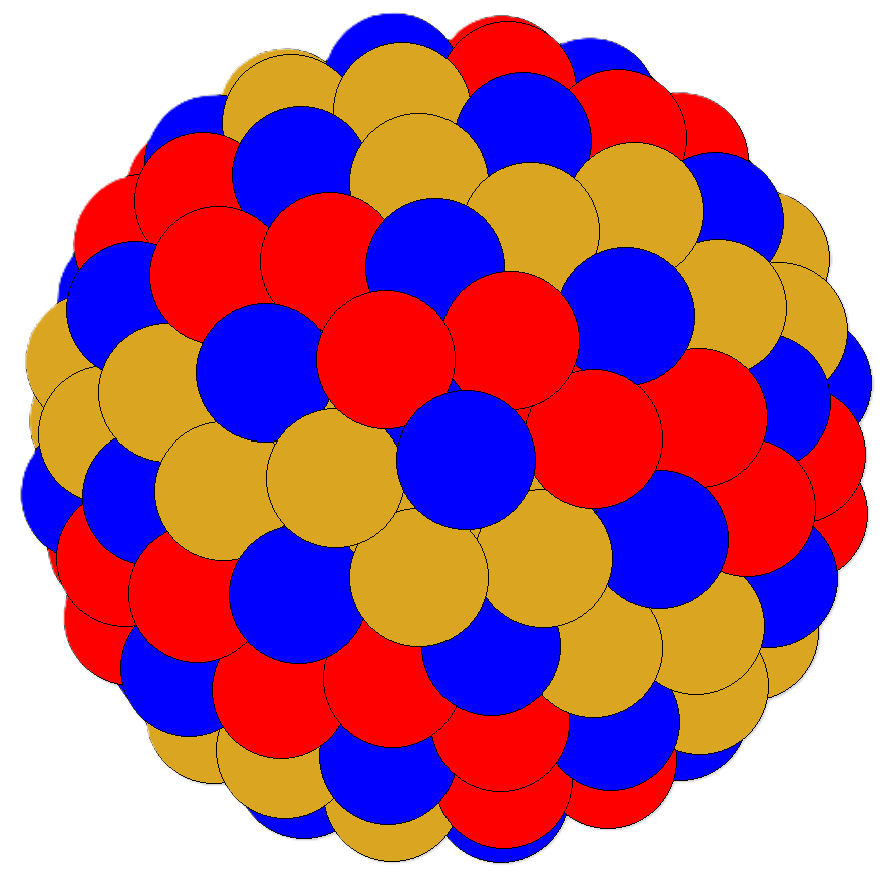}\\
    \# 3 (9)
  \end{minipage}\hfill
  \begin{minipage}[t]{0.24\textwidth}
    \centering\includegraphics[height=3cm]{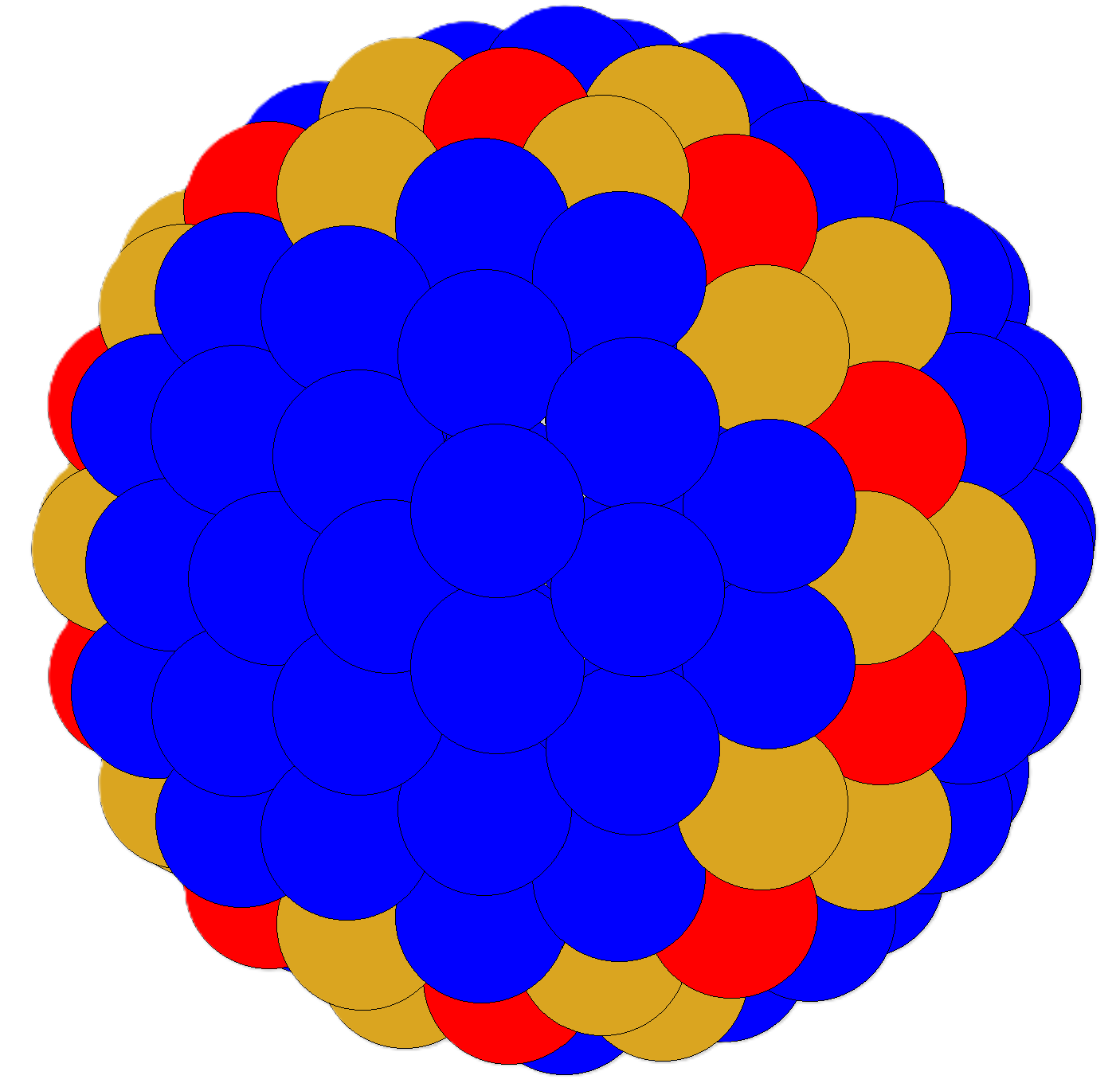}\\
    \# 4 (11)
  \end{minipage}\\ \vspace{1em}
  
  \begin{minipage}[t]{0.24\textwidth}
    \centering\includegraphics[height=3cm]{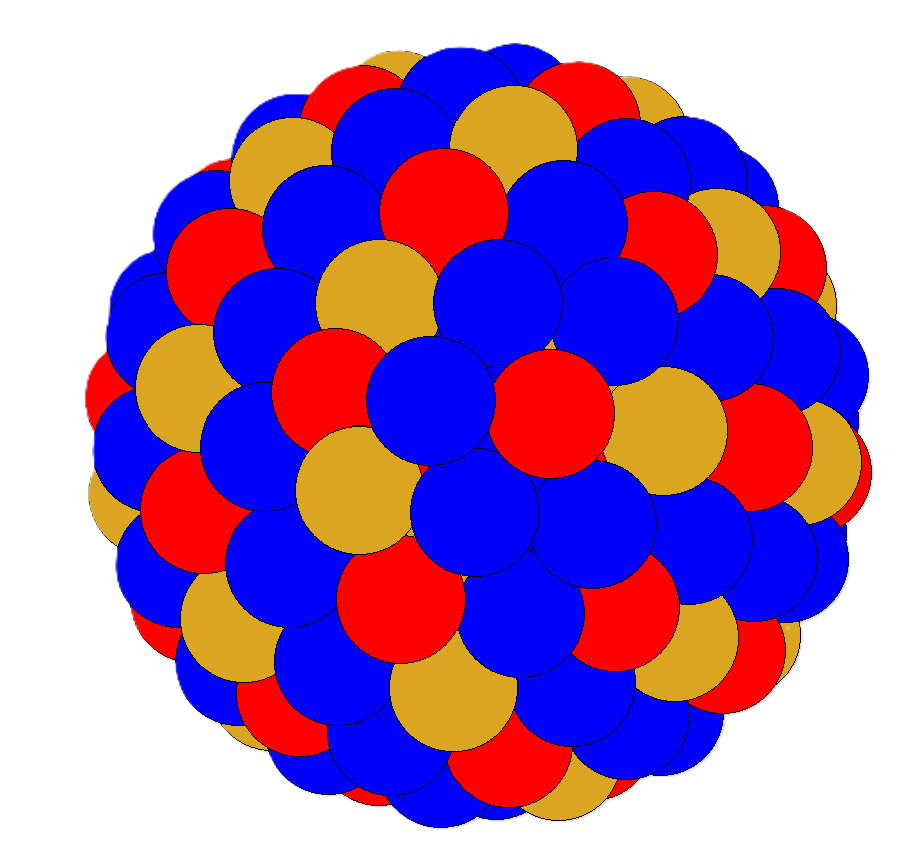}\\
    \# 5 (4,33)
  \end{minipage}\hfill
  \begin{minipage}[t]{0.24\textwidth}
    \centering\includegraphics[height=3cm]{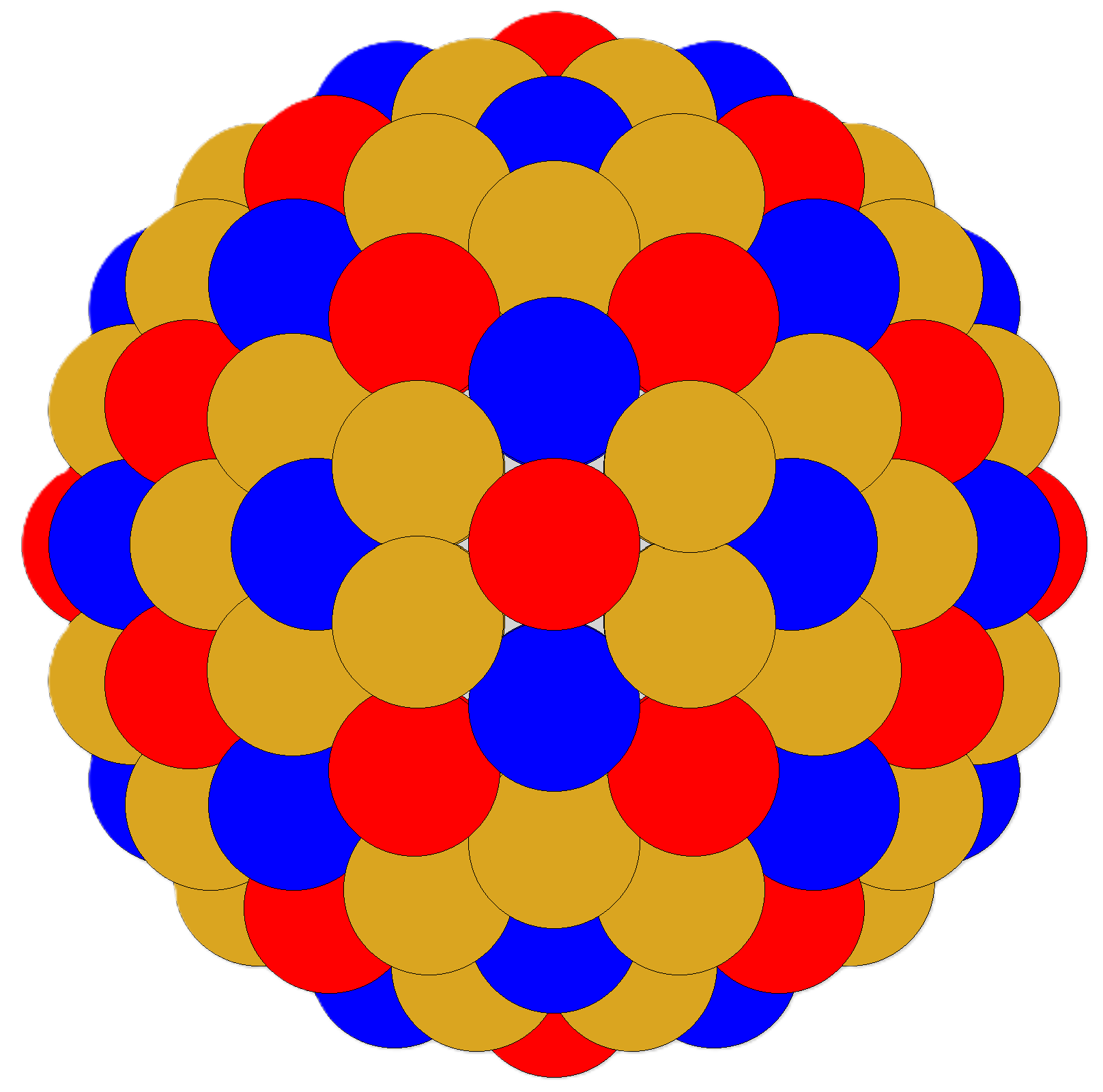}\\
    \# 6 (18)
  \end{minipage}\hfill
  \begin{minipage}[t]{0.24\textwidth}
    \centering\includegraphics[height=3cm]{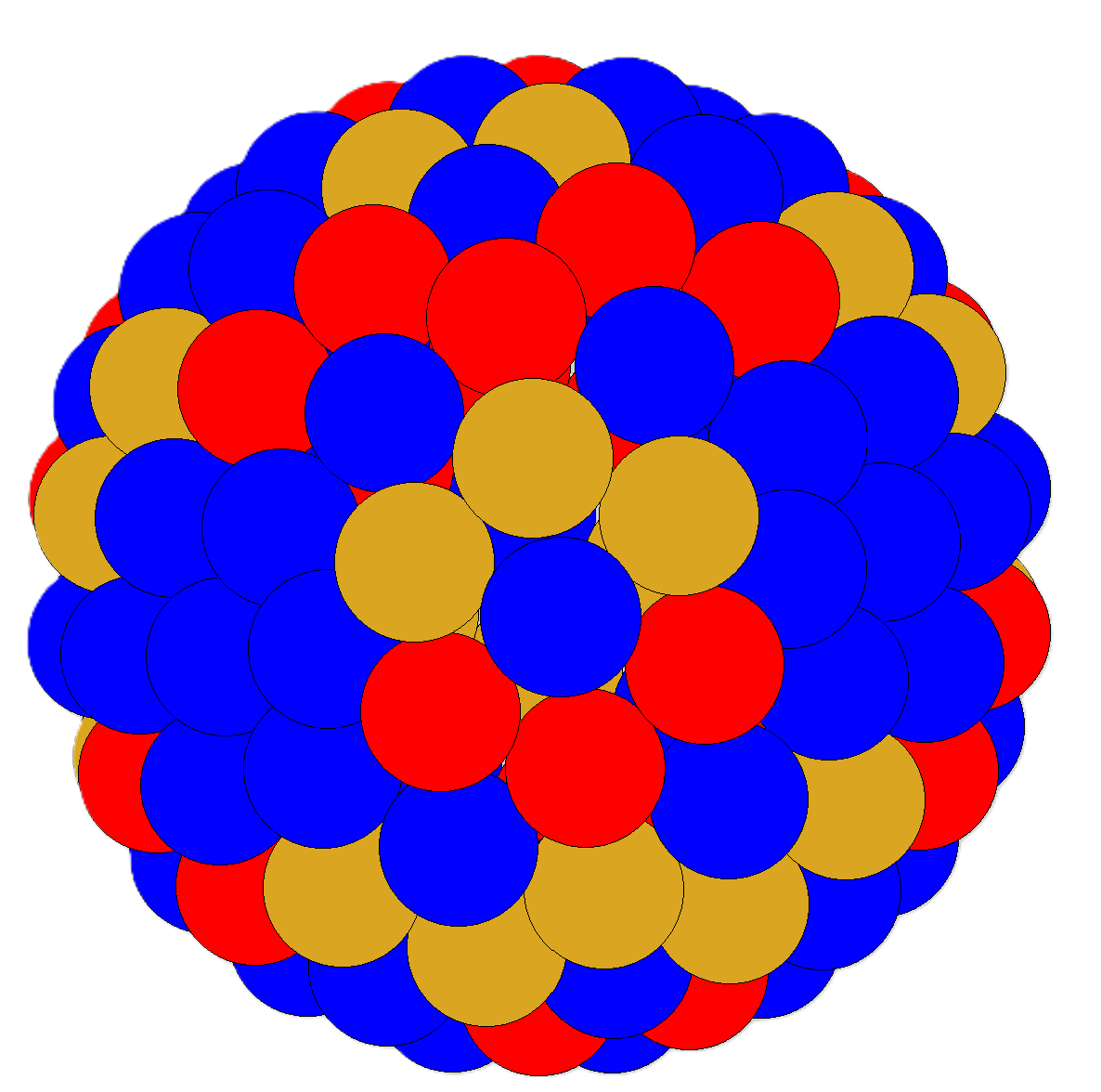}\\
    \# 7 (28)
  \end{minipage}\hfill
  \begin{minipage}[t]{0.24\textwidth}
    \centering\includegraphics[height=3cm]{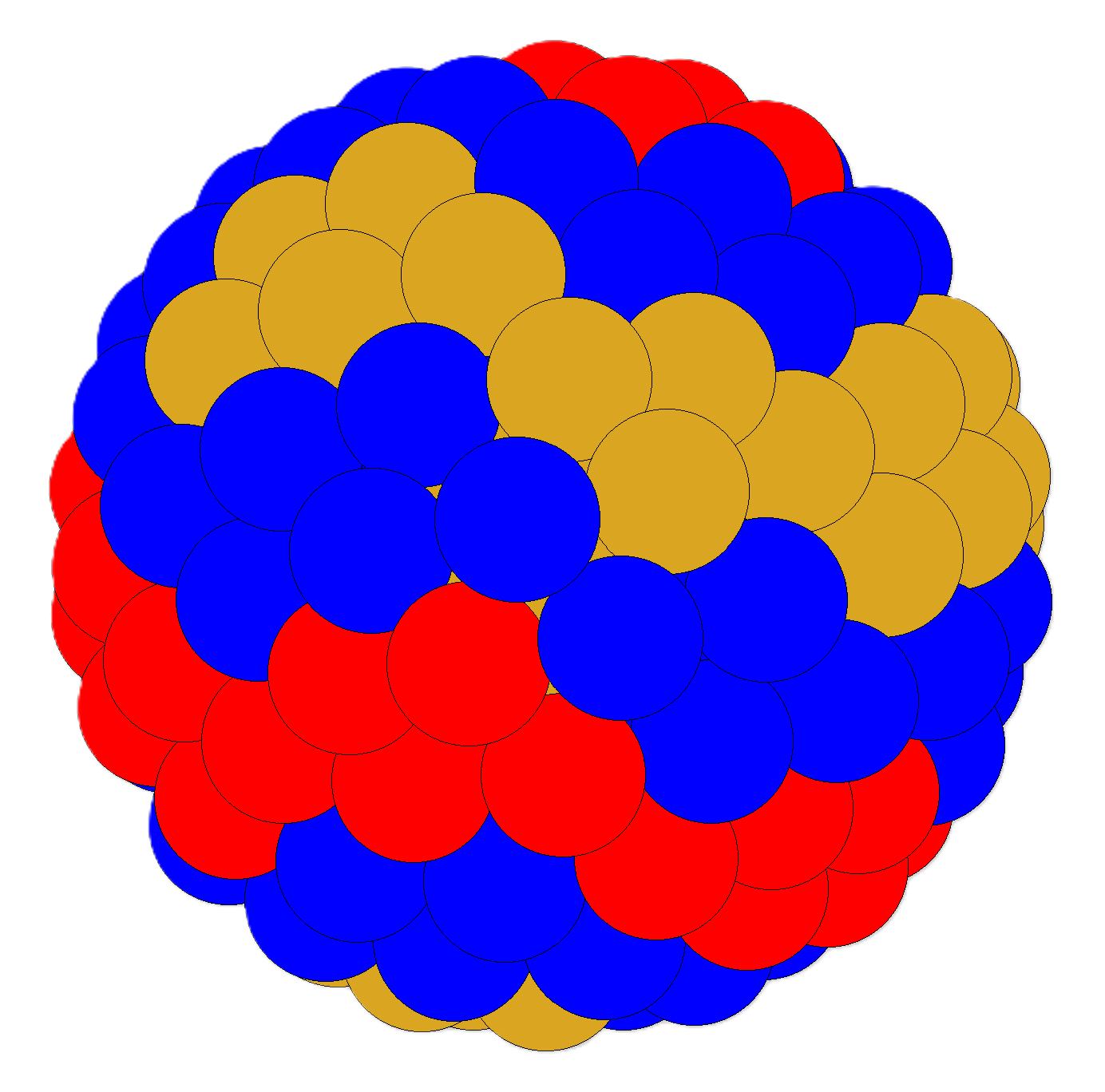}\\
    \# 8 (36)
  \end{minipage}\\ \vspace{1em}
  
  \begin{minipage}[t]{0.24\textwidth}
    \centering\includegraphics[height=3cm]{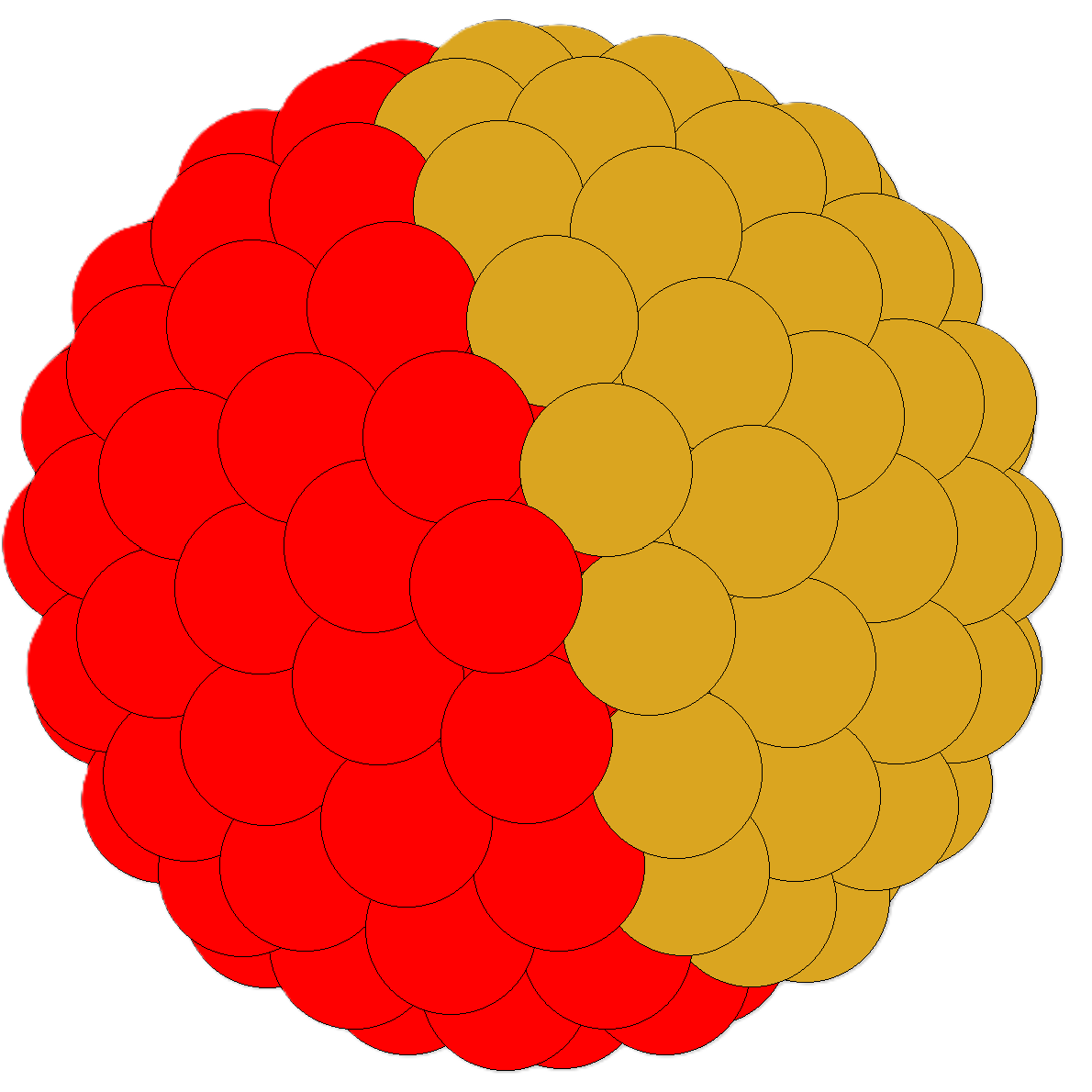}\\
    \# 9 (36)
  \end{minipage}\hfill
  \begin{minipage}[t]{0.24\textwidth}
    \centering\includegraphics[height=3cm]{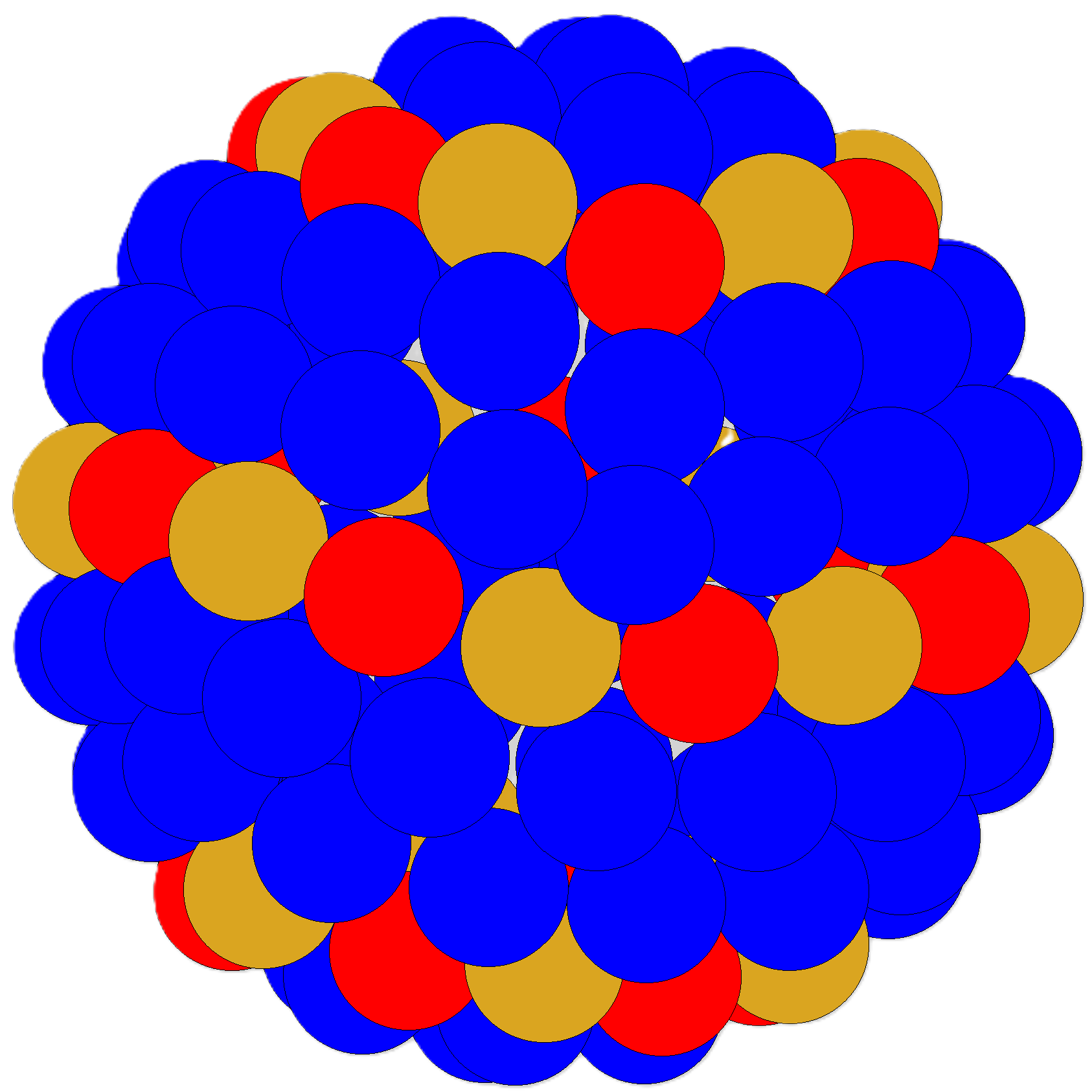}\\
    \# 10 (26)
  \end{minipage}\hfill
  \begin{minipage}[t]{0.24\textwidth}
    \centering\includegraphics[height=3cm]{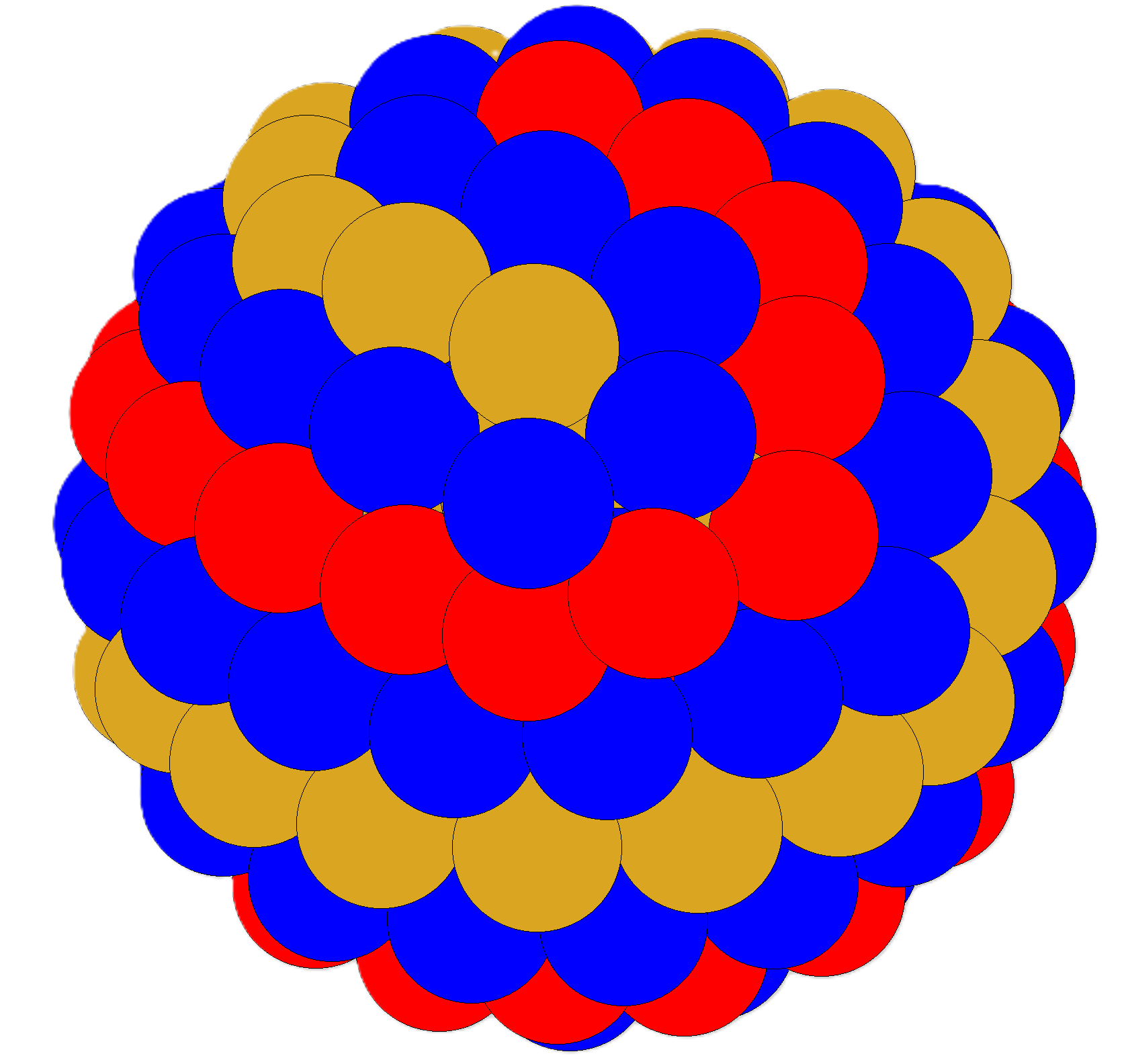}\\
    \# 11 (10)
  \end{minipage}\hfill
  \begin{minipage}[t]{0.24\textwidth}
    \centering\includegraphics[height=3cm]{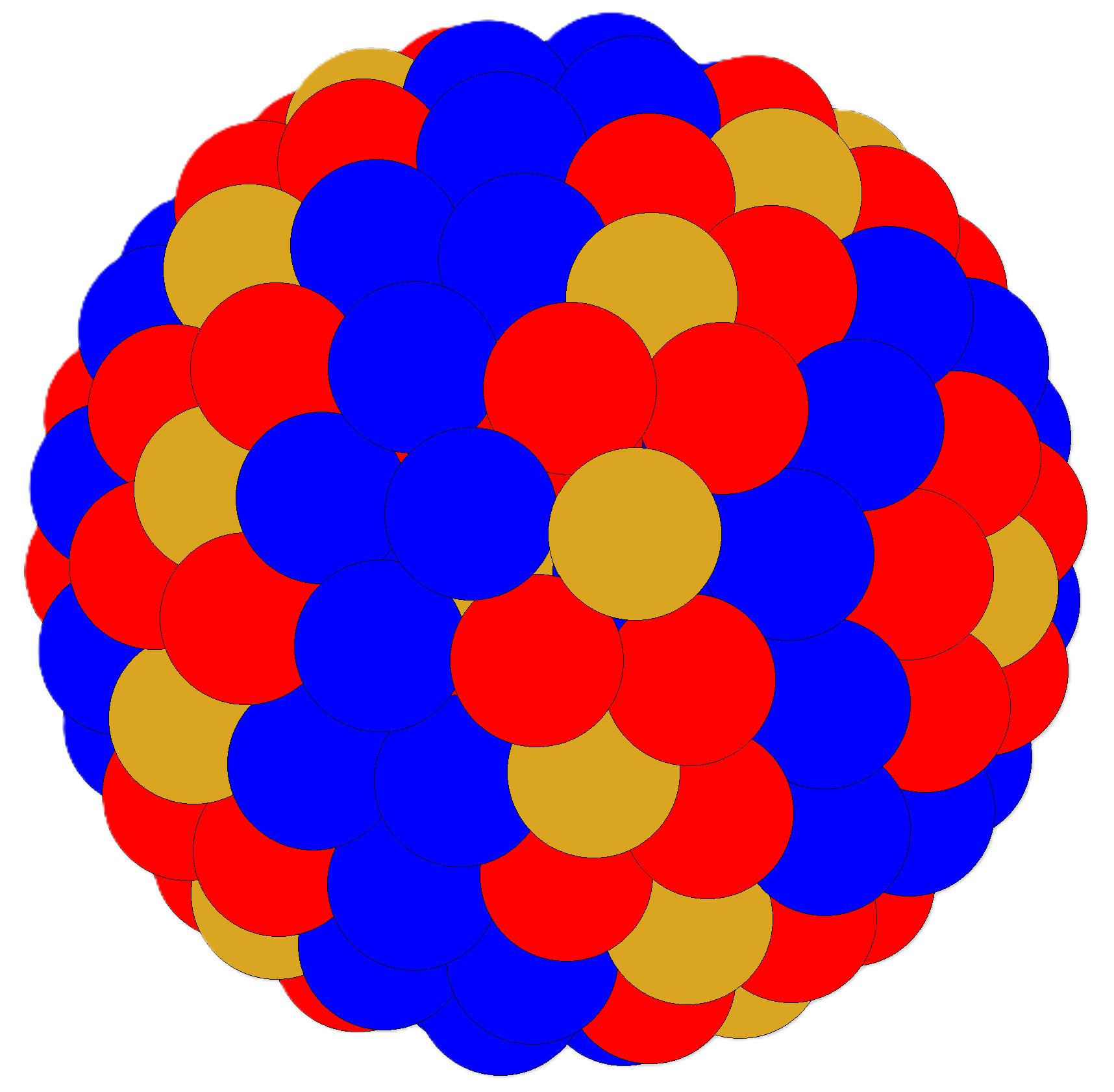}\\
    \# 12 (11)
  \end{minipage}\\ \vspace{1em}

    \begin{minipage}[t]{0.24\textwidth}
    \centering\includegraphics[height=3cm]{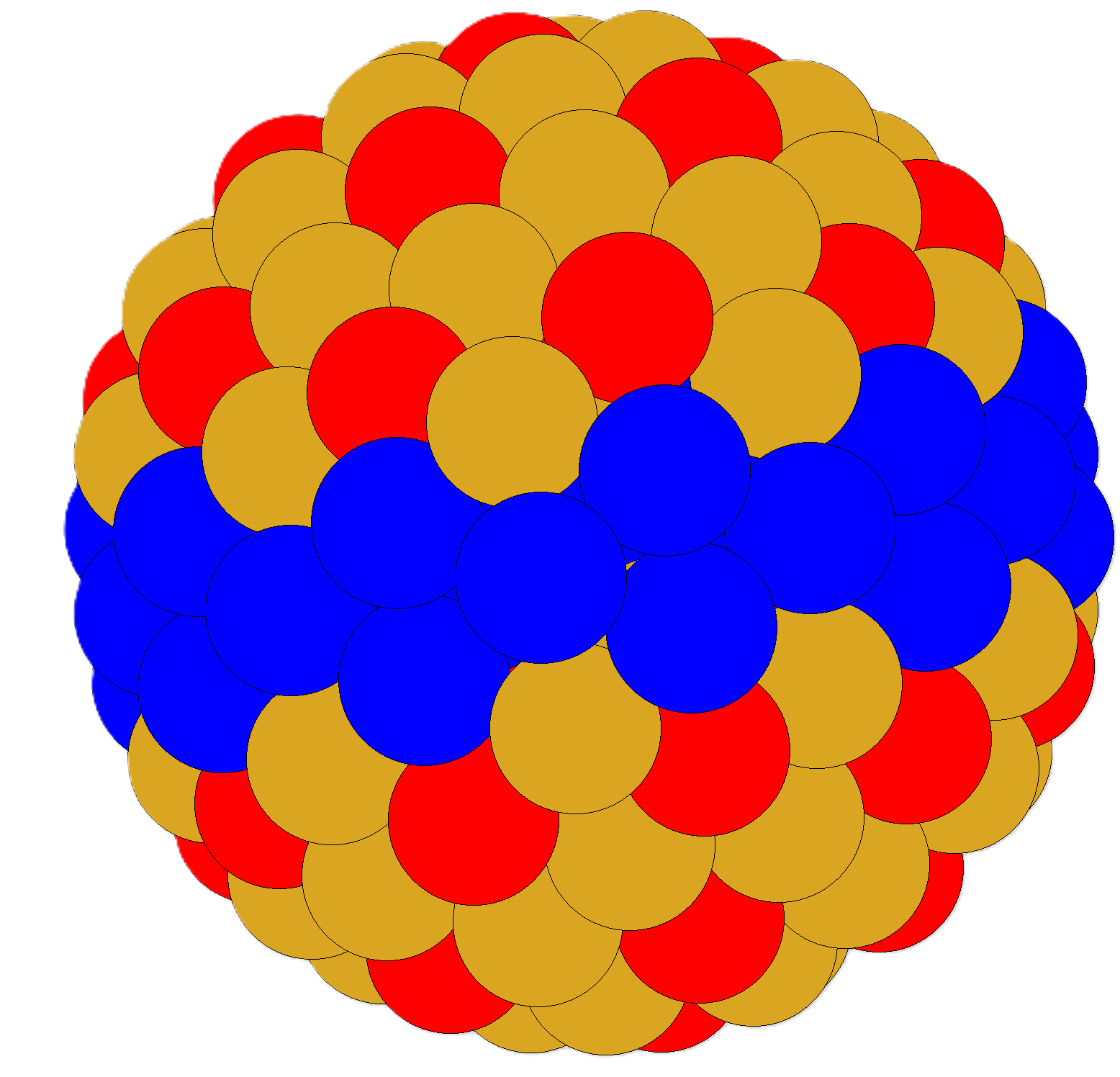}\\
    \# 13 (11)
  \end{minipage}\hfill
      \begin{minipage}[t]{0.24\textwidth}
    \centering\includegraphics[height=3cm]{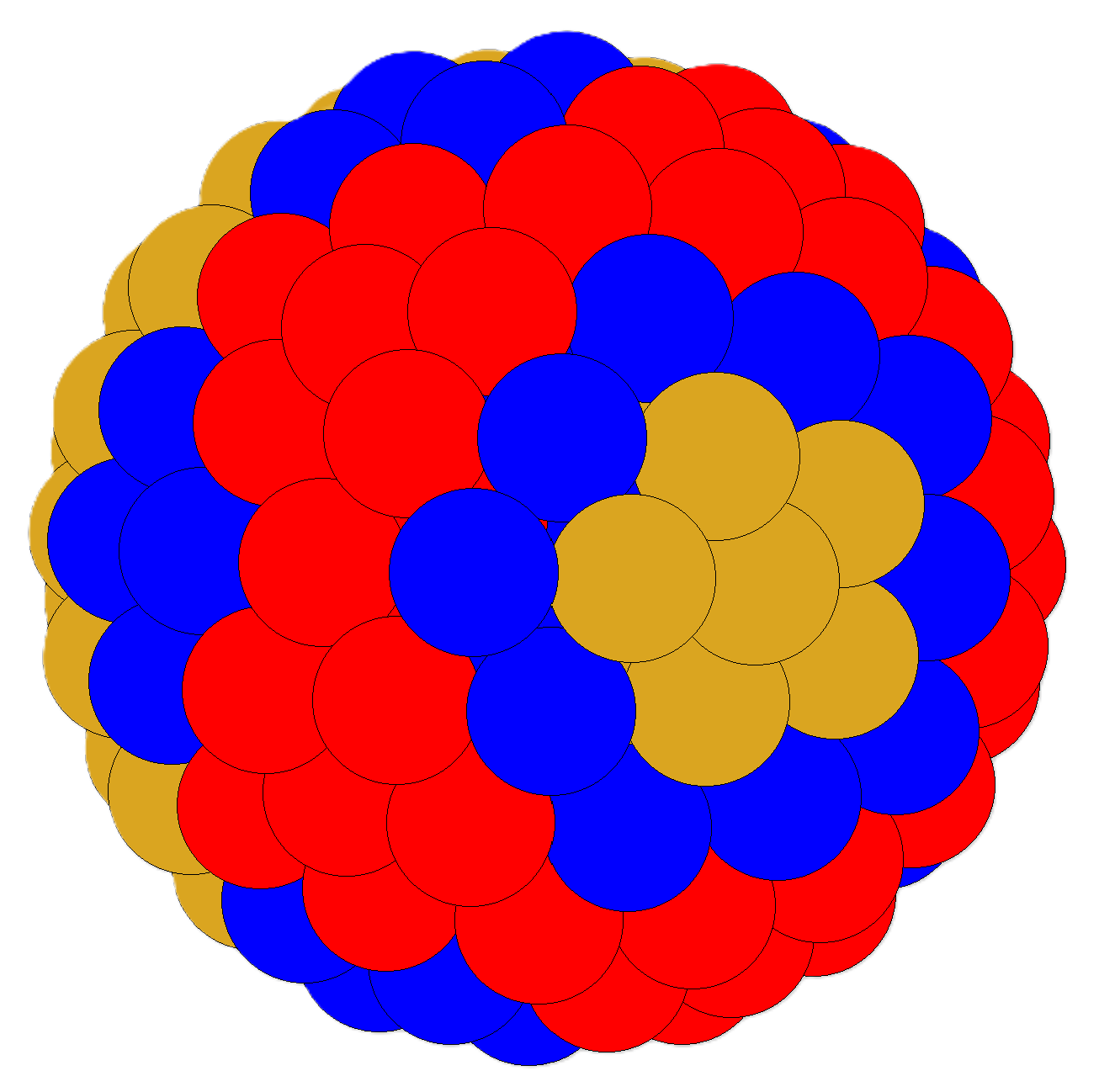}\\
    \# 14 (36)
  \end{minipage}\hfill
        \begin{minipage}[t]{0.24\textwidth}
    \centering\includegraphics[height=3cm]{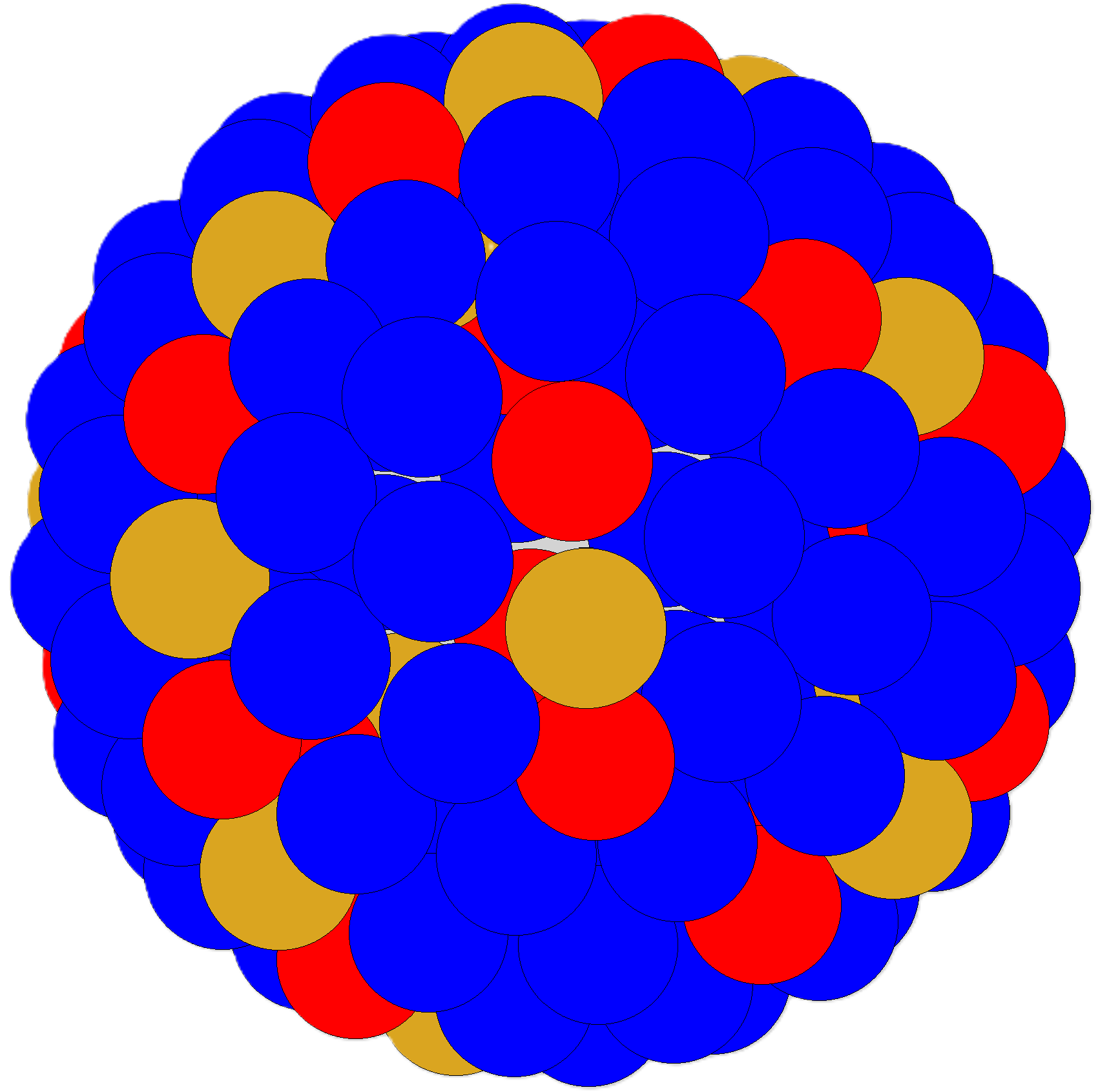}\\
    \# 15 (26)
  \end{minipage}\hfill
          \begin{minipage}[t]{0.24\textwidth}
    \centering\includegraphics[height=3cm]{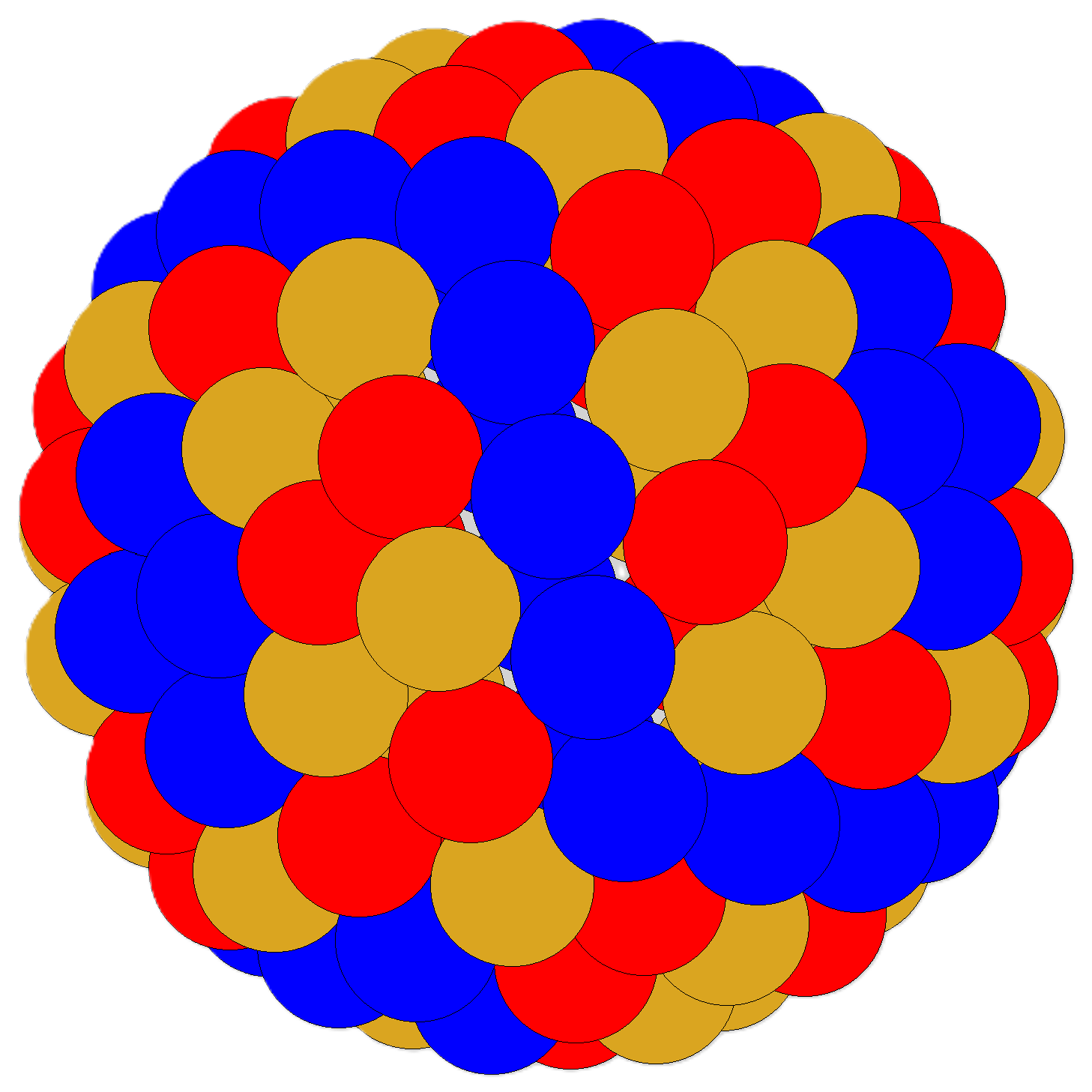}\\
    \# 16 (26)
  \end{minipage}
  \caption{Patterns of a lattice-gas mixture on the spherical grid.
Some of these patterns are commented in the text}
  \label{gallery4}
\end{figure*}

For stripes \# 8, which is a paradigmatic example of two-component stripes, the energy per particle is $3u_{1,1}^{(3)}+u_{1,2}^{(1)}+u_{1,2}^{(2)}$.
In this expression, first- and second-shell like-particle interactions are not present.
Therefore, if $u_{11}^{(3)}$ is zero or negative, a repulsion at shorter distances inevitably suppresses many configurations competing against stripes, which are thus encouraged to appear.
Indeed, stripes \# 8 are observed for $U_{11}=(1,1,0)$ and $U_{12}=(-1,0,0)$ (case 22).
They are also present as $N=72$ minimum-energy configurations for the interactions $U_{11}=(1,1,0)$, $U_{12}=(-1,-1,0)$ and $U_{11}=(1,0,-1)$, $U_{12}=(-1,0,0)$ (cases 23 and 33), even though in both cases stripes lack sufficient stability to produce a plateau in $N(\mu)$.
The previous ones are just a few examples where the emergence of stripes can be accounted for based on the interaction features.
We note that stripes are even promoted in cases, like case 33, where a Lennard-Jones-type interaction between like particles is combined with a SW interaction (hence, Lennard-Jones-type too) between particles of different species.
The same arguments used above also serve to explain the onset of curved stripes on the spherical grid.
Actually, a more or less similar reasoning applies for all stripe patterns in Fig.~\ref{gallery1}, although it is clear that in a two-dimensional continuum model of binary mixture the interactions cannot be exactly programmed {\it a priori} to yield a particular stripe pattern, and a further element of uncertainty remains, related to the possible promotion of non-stripe structures that are only slightly suboptimal on the lattice.
For instance, a close alternation of colors within the same stripe would be a likely eventuality when the interaction between unlike particles is attractive at first-neighbor distance, as in cases 22 and 26.
However, case 4 indicates that two-color stripes may also appear when the unlike interaction is purely hard-core.
Another example is case 18, where two-color stripes are instead promoted by a third-neighbor attraction between unlike particles.
We have previously mentioned that, in binary mixtures, the notion of stripes extends even beyond the paradigm of translationally invariant modulated structures.
Examples of stripe patterns holding some degree of irregularity at $T=0$ are shown in Fig.~\ref{gallery2}.
In the thermodynamic limit, these structures correspond to low-temperature phases that can rightfully be dubbed ``stripe liquids''.
It is worth noting that the interactions giving rise to irregular stripes are anything but few;
furthermore, for case 27 we observe (at different chemical potentials) two distinct kinds of irregular stripes.
As illustrated in Fig.~\ref{gallery3}, the level of complexity may actually be higher.
On each row of the figure, we show ground-state configurations (either regular or irregular) sharing the same density and energy;
all these structures are coexisting at $T=0$, in that the mixture can switch from one to the other at zero free-energy cost.
Among these patterns some are stripes, while in others worm-like structures occur together with clusters.

\begin{table*}
\caption{\label{tab:prova}
Self-assembled structures of non-symmetric mixtures on the spherical grid.
In all the listed cases, species-1 particles have zero off-core interaction, i.e., $U_{11}=(0,0,0)$.}
\begin{tabular}{|l|c|c|c|}
\hline
\textbf{Case}&\textbf{$U_{\rm 22}$}& \textbf{$U_{\rm 12}$} & \textbf{Low-temperature structures} \\
\hline
$41*$&$(-1,0,0)$ & $(-1,0,0)$ &$N=122$: Isolated hard spheres in a sea of LJ-type particles\\
\hline
$42$&$(-1,0,0)$ & $(-1,-1,0)$ & Worms and clusters of hard sphere in a sea of LJ- particles\\
\hline
$43$&$(1,-1,0)$ & $(-1,0,0)$ &  $N=122$: Crystal-like structure\\
\hline
$44*$&$(1,-1,0)$ & $(-1,-1,0)$ & $N=122$: Stripes\\
\hline
$45*$&$(1,0,-1)$&$(-1,0,0)$& Small $N$: Crystal of LJ-type particles.
$N=122$: Stripes\\
\hline
$46*$&$(1,0,-1)$&$(-1,-1,0)$& Large $N$: Stripes\\
\hline
$47*$&$(-1,2,1)$ & $(-1,0,0)$ & $N=122$: Cluster crystal of SALR particles in a sea of hard spheres\\
\hline
$48*$&$(-1,2,1)$ & $(-1,-1,0)$ & $N=122$: Worms and clusters of SALR particles in a sea of hard spheres\\
\hline
$49*$&$(-1,0,1)$&$(-1,0,0)$& $N=122$: Defective crystal-like pattern\\
\hline
$50*$&$(-1,0,1)$ & $(-1,-1,0)$ & $N=122$: Stripe-like pattern\\
\hline
\end{tabular}
\end{table*}

Finally, we discuss the peculiarities of self-assembly in mixtures of lattice gases on a spherical grid.
We remark that any polyhedral grid can in principle be fabricated by DNA origami~\cite{Benson2015}, and then used as a template for the adsorption of two species of particles.
The spherical topology heavily affects the order exhibited by the mixture at low temperature, inducing frustration and ruling out the possibility of proper crystalline or stripe structures.
However, these arrangements are not completely washed out on the sphere, but just ``deformed'':
for each crystalline or stripe structure on the triangular lattice, a similar pattern exists on the spherical grid (see many examples in the Supplementary Material).
In addition to these patterns, other curved patterns exist, having no counterpart on the triangular lattice, which could not simply have been predicted from an inspection of the Hamiltonian.
We have collected in Fig.~\ref{gallery4} a few notable cases, chosen among those showing some built-in symmetry that might be exploited to create synthetic ``particles'' with specific functionalities.
Besides the practical interest, what is worth noting in Fig.~\ref{gallery4} is the extreme complexity and variability of the self-assembly behavior induced, on a geodesic grid, by very basic interaction rules.
For example, in pattern \# 1 particles form pentagonal rings that are overall arranged in a rigid network, with holes at the vertices of a regular icosahedron.
While a similar structure exists on the triangular grid, see cases 19 and 21 in the Supplementary Material, it is only on the spherical grid that this decoration can be used to coordinate twelve larger particles (in the network holes), alternately bonded along each ring with species-1 and species-2 particles.
More numerous adsorbing centers are seen in patterns \# 3 and \# 6, placed at the vertices of a geodesic polyhedron (the pentakis icosidodecahedron and the pentakis dodecahedron, respectively).
In pattern \# 4 particles are arranged in a ring dividing the sphere into two regions of different sizes.
The (center of the) smaller hole can serve as adsorbing site for a foreign particle of suitable diameter, this way producing an asymmetric dimer.
In patterns \# 5 and \# 10, particles of alternating species are arranged in three parallel rings;
we can imagine to fill the voids between the rings with particles of a third species, so as to form a thick ribbon wrapped around the sphere.
A similar ribbon could be formed, starting from pattern \# 12 or \# 13, with foreign particles adsorbed in the empty channel along the equator.
As a last example, pattern \# 9 represents a concrete realization of a Janus particle~\cite{Walther2013} (with a slight preponderance of one ``color'' over the other).

In closing this Section, we briefly mention the existence, in addition to stripe patterns, of a wide variety of spontaneously self-assembled orderly structures akin to crystals and cluster crystals, both on the triangular and on the spherical grid.
However, a thorough examination of these arrangements is beyond the scope of this paper;
the interested reader can find a catalog of them in the Supplementary Material.
\section{Results: Non-symmetric mixtures}

Next, we consider a few cases of mixtures of two non-equivalent species on the spherical grid, while still keeping $\mu_1=\mu_2$.
In particular, species-1 particles are assumed to be purely hard-core.
We allow for a SW interaction between the species, $U_{12}$, of two possible ranges.
The results are briefly summarized in Table II, while the patterns are illustrated graphically in the Supplementary Material (hard particles in red).
When the species are different, the constraint $\mu_1=\mu_2$ is not necessarily synonym of near-equimolarity.
For example, when $U_{22}=(-1,0,0)$ the $N=122$ microstates of minimum energy are rich in species-2 particles, while species 1 is only marginally present and its structure is disordered.
The compositional disequilibrium is reduced when the range $U_{12}$ is longer.
Notably, in the latter case we observe the formation of clusters of particles that are otherwise non-interacting.
When $U_{22}=(1,-1,0)$ (cases \# 43 and \# 44), the $N=122$ pattern depends on the range of $U_{12}$:
crystal-like for $U_{12}=(-1,0,0)$ and stripe-like for $U_{12}=(-1,-1,0)$.
In the former case, species-2 particles form a pentakis icosidodecahedron, while the other particles sit at the vertices of a chamfered dodecahedron.
By shifting the $U_{22}$ well to third-neighbor distance (cases \# 45 and \# 46), the $N=122$ pattern is stripe-like for both instances of $U_{12}$.
The $N=122$ pattern of a mixture of hard-core particles and SALR particles depends on the height of the repulsive barrier (cases \# 47-\# 50):
cluster-crystalline for a higher barrier and stripe-like for a lower barrier --- but only if $U_{22}=(-1,-1,0)$.
\section{Results: Equimolar mixtures}

Finally, we have considered a few cases of symmetric mixtures on the spherical grid where the only admissible microstates are those with an equal number of species-1 and species-2 particles.
The most relevant patterns of these strictly-equimolar mixtures are shown in the Supplementary Material.
With the obvious exception of the first two cases, where the equimolarity constraint prevents the expulsion of one species from the grid, only marginal differences are detected with the cases where $\mu_1=\mu_2$ --- a sign that the latter condition actually results in near-equimolarity.
\section{Conclusions}

The self-assembly of impenetrable classical particles on a finite grid retains the richness and complexity of pattern formation in continuous space.
This is vividly demonstrated by the behavior of near-equimolar mixtures of two particle species on the triangular grid and on a geodesic grid of 122 sites.
Given any particle interaction up to third-shell sites, the exact equilibrium states of the mixture can be worked out, with moderate effort, using Wang-Landau sampling.
By this method, the interaction space of the model can be swept very efficiently and general self-assembly trends can be found out.
As a hosting space for particles, a grid composed of a hundred sites is large enough to expect that the mixture behaves like a thermodynamic system.
Indeed, the average number $N$ of particles (i.e., occupied sites) exhibits, as a function of the chemical potential $\mu$, a series of plateaus at low temperature (``patterns''), akin to those observed in a many-body system undergoing a sequence of first-order phase transitions.
By reviewing dozens of interaction profiles, we have identified a wide variety of patterns that are robust to heating --- since the relative plateaus in $N(\mu)$ are almost unaffected by a raise of temperature, e.g., from $0.05$ to $0.5$ (see, for instance, cases 9, 18, 25, and 26 in the Supplementary Material).
Furthermore, self-assembled structures on the spherical grid look locally similar to those emerging on the plane.
Among the emergent structures, special attention is paid to stripe patterns, occurring in many shapes and for many different combinations of like and unlike interactions.
We have found instances of both one-component and two-component stripes, and cases where stripes stay irregular even at zero temperature.
Moreover, in some phases stripes are not straight but wavy, although we were not able to find any straightforward relationship between the shape of interactions and that of stripes.
The only correlation we have devised concerns the thickness of stripes:
the more extended the attraction between unlike particles, the thicker the stripes.
However, some other considerations have nonetheless emerged from our study.
For example, stripes are energetically favored over more symmetric (or disordered) configurations whenever a short-range repulsion between particles of same species is accompanied by a longer-range attraction between unlike particles --- the longer the range of attraction, the thicker the stripes.
When like interactions are longer-range repulsive, stripes even occur with no attraction between unlike particles (since all competitor configurations have a higher free-energy cost).
We document the existence of stripes in several other situations, involving either a Lennard-Jones-type or a SALR interaction between like particles, combined with an attraction between unlike particles (a comprehensive list of stripe-forming interactions is reported in the Tables and in the Supplementary Material).
The takeaway message is that the emergence of stripe order at low temperature does not require an extremely fine tuning of interactions;
as a result, stripes will be more easily developed in binary mixtures than imagined so far.
If the cooperative behavior is already rich on the triangular lattice, the degree of structural complexity exhibited by a lattice-gas mixture on the spherical grid is even higher.
Far from being an exotic construct, polyhedral grids can be fabricated, on a submicron scale, by DNA origami~\cite{Benson2015}.
Regular polyhedral templates can also be realized with block copolymers, as witnessed by, e.g., TEM images of PS-b-PI nanoparticles~\cite{Kagawa2025}.
When exposed to a gaseous mixture of two distinct chemical substances (e.g., two species of colloidal particles), the template can be ``decorated'' by just letting the system attain thermodynamic equilibrium at low temperature.
By this way, patchy particles of various types can be engineered.
Clearly, the type and regularity of the embellishment depends on the effective interaction between the particles adsorbed over the template.
Despite control of the mutual forces between the latter particles being in practice limited, the relative abundance of stripe-forming interactions suggests that at least some of these surface ornaments will be stripe patterns.
More generally speaking, knowing in advance the interactions producing the desired self-assembly of adsorbed particles can be useful in the design of functional materials with customized surface properties.

\section{Supplementary Material}
Graphical information relative to all lattice-gas mixtures analyzed in our paper can be accessed in the supplementary material.
\bibliography{paper}
\end{document}